\documentclass[twocolumn]{aastex631}

\usepackage{newtxtext,newtxmath}

\usepackage[T1]{fontenc}
\usepackage{ae,aecompl,ulem,color}

\usepackage[utf8]{inputenc}
\usepackage{graphicx}
\usepackage{amssymb}
\usepackage{amsmath}
\usepackage{xcolor}
\usepackage{natbib}
\usepackage{ulem}

\newcommand{\Ksc}{\epsilon_\mathrm{e}/\epsilon_\mathrm{_B}} 
\newcommand{\epse}{\epsilon_\mathrm{e}} 
\newcommand{\epsB}{\epsilon_\mathrm{_B}} 
\newcommand{\SPN}{SPN98 coefficients} 
\newcommand{\new}{effective coefficients} 

\shorttitle{GRB~190114C afterglow parameters}
\shortauthors{Derishev \& Piran}

\begin{document}

\title[GRB~190114C afterglow parameters]{GRB afterglow parameters in the era of TeV  observations: the case of GRB~190114C}
\author[0000-0002-6761-5515]{Evgeny  Derishev}
\affil{Institute of Applied Physics RAS, 46 Ulyanov st, 603950 
Nizhny Novgorod, Russia}
\author[0000-0002-7964-5420]{Tsvi Piran}
\affiliation{Racah Institute of Physics 
The Hebrew University of
			Jerusalem, Jerusalem 91904, Israel}

\begin{abstract}
The afterglow of GRB~190114C has been observed at {60-1200~s} after the burst in the {sub-TeV range} by the {MAGIC Cherenkov} telescope.
The  
simultaneous observations in X-ray range, which is presumed to be of synchrotron origin, and in sub-TeV range, where the emission is presumed to be  inverse Compton,  provide new stringent constraints on the conditions within the emitting regions {and their evolution in time}. While the additional data contain a lot of new information, it turns out that fitting both the X-ray and the TeV emission,  is much more complicated than what was originally anticipated. 
{We find that optical flux measurements provide important complementary information that in combination with TeV measurements breaks degeneracy in the parameter space.}
We present here a  numerical fit  to the multi-wavelength observed spectrum  using a new code that calculates the single-zone synchrotron including self-Compton emission taking into account the exact Klein-Nishina cross-sections as well as pair production via  absorption of the high-energy photons inside the emitting zone and the emission from the resulting secondary pairs. 
We also present a revised set of single zone parameters  and a method for fitting the data to the observations.  Our  model for GRB~190114C that fits all the observations, from the optical data point to the sub-TeV range,  suggests that it is in the fast cooling regime. The inferred parameters for  observations {at two separate moments of time} show significant deviations from some of the common expectations in afterglow modeling but are all consistent with the predictions of the {pair-balance} model. 
\end{abstract}

\keywords{Gamma-ray bursts; Non-thermal radiation sources; Relativistic jets; Shocks}

\section{Introduction}

Gamma-Ray Bursts (GRBs),  both short (that last a fraction of a second) and long (up to hundreds of seconds) ones, are followed by long lasting multiwavelength afterglows that  have been observed in
X-rays, IR/optical/UV and Radio. The observations have been interpreted as synchrotron emission arising from a relativistic blast wave propagating into {interstellar medium (ISM) or stellar wind} \citep{MeszarosRees97,AfterglowModelling}.  The blast wave generates magnetic fields and accelerates electrons that  produce this radiation. While numerous modifications and extensions  to this model have been suggested, and while some observations in various burst don't conform to the simplest version of this model, 
its predictions are in rough agreement with the observations.

The basic afterglow model 
\citep{AfterglowModelling,GranotSari} is a one-zone model {where energetic electrons downstream of a relativistic shock emit synchrotron radiation in shock-generated magnetic field.} The electrons'  and magnetic energies are related to the total energy behind the shock using constant equipartition parameters. Depending on parameters,  the system can be in fast or slow cooling (in the former case most of the electrons'  energy is emitted within a dynamical time scale, in the latter it is not) or in an adiabatic or radiative evolution (in the former case the blast wave energy is much larger than the radiated energy, in the latter, which can take place only in fast cooling, a significant fraction of the blast wave energy is radiated in a dynamical time scale).

The model results in closure relations between the spectral indices in different spectral {ranges} and the temporal behaviour. While the model has been giving a good overall fit to the observations, there are  well known deviations from  its  basic version   (plateaus, sharp decline, flares and others). In spite of those, it has  been extensively used  to infer the conditions within the afterglows' emitting regions  and other global parameters, such as the total kinetic energy of the blast wave or the density distribution of the circum-burst matter \citep[see e.g.][and many others]{WijersGalama1999,PanaitescuKumar2000,PanaitescKumar2002,vanEerte+2012,Nava+13,Beniamini2015,Ghirlanda+2018,Aksulu2020}. 

The same electrons that produce synchrotron emission will also produce inverse Compton (IC) radiation  \citep[see e.g.][and many others for discussion in the context of GRBs and their afterglow] {SariNarayanPiran96,PanaitescuMeszaros98,SariEsin2001,PeerWaxman2005,FanPiranNarayan2008,AndoNakarSari2008,NakarAndoSari2009}.  
Furthermore, \cite{DerishevKocharovskyKocharovsky2001} have shown  that typical GRB synchrotron emission must be accompanied in the TeV range by an IC component, whose power is at least $\sim 10$ per cent of the synchrotron power. Thus, observations  of  the high energy component would add significant new constraints on the emission mechanism and will enable a clearer determination of the conditions within the emitting region, such as the ratio of electrons to magnetic energy densities as well as  the typical Lorentz factors of the emitting electrons. 
However, it turns out that a determination of expected {spectrum}  when {Klein-Nishina (KN) corrections} take place
is rather complicated.   
{Analytic description of how IC losses in KN regime influence the synchrotron part of SED was given by \cite{DerishevKocharovskyKocharovsky2001} and then in greater detail and in application to the afterglow evolution by \cite{NakarAndoSari2009} } 
\citep[see also][]{
KNeffects_Fan,KNeffects_DaigneBosnjakDubus,KNeffects_BarniolDuranBosnjakKumar,KNeffects_WangHeLi,KNeffects_Lemoine,KNeffects_TakOmodeiUhm}. 
In fact as we show below the situation is even more complicated than the analytic expectation when one attempts a full numerical solution.  

In this paper we describe a method to estimate the conditions within the emitting region given a set of observations at low energies (X-rays) corresponding to synchrotron and at high energies (TeV) corresponding to  self-Compton (SSC). 
We characterize these conditions using five variables: $\Gamma$, the Lorentz factor of the shock {front}, $\gamma_\mathrm{m}$ (or $\gamma_\mathrm{b}$, which we define later) and $p$, the characteristic Lorentz factor of the emitting electrons and the slope of  the  high energy {tail of the injection function}, $B$, the magnetic field in the emitting region, and $\Ksc$, the ratio of the electrons' to magnetic field energy density. 
The method is based on a new single zone code \cite{newcode} that includes for a given injection spectrum all relevant radiation processes --- synchrotron and SSC within the KN regime,   pair creation by high-energy photons (as a result of their absorption inside the emitting zone) and the  radiation from secondary pairs produced. 
We apply this method to the observed data of GRB~190114C \citep{MAGIC_Nature_obs} and compare our results to other modeling of this event \citep{MAGIC_Nature_fit,Wang_etal_SED,Asano_etal_SED}. 

GRB~190114C was observed by MAGIC in the sub-TeV {range} $\approx 60 \div 1200$ seconds after the trigger. We find that while details of our fit are somewhat different from our earlier results \citep{DerishevPiran2019}, that were analytic estimations based only on the preliminary GCN data, the basic features of fast cooling and a rather large value for the magnetic equipartition parameter ($\epsB$) hold. 

Our results agree with the basic predictions of the pair balance model \citep{PairBalance}.
{ This model  makes {specific predictions} about the average Lorentz factor of {injected} electrons {and the ratio of synchrotron and IC powers}. {It's framework} includes an ``accelerator'', which supplies energy to radiating particles, and an ``emitter'', which {transfers} energy from the particles to synchrotron and IC radiation. The interaction between the two is in the form of two-photon pair production due to internal absorption of high-energy IC photons by low-energy target photons (of synchrotron origin). 
A key feature of the pair balance model is a strong shock modification, which smears out the discontinuity.  
The latter completely disappears if the power of high-energy (IC) radiation absorbed inside the shock is $\sim 15$ per cent of the shock kinetic power, switching off 
runaway acceleration of secondary pairs \citep{PairBalance}. 
The need to balance the accelerator’s power by pair loading results in the requirement that the emitting zone is not entirely transparent to its own IC radiation.
It also drives the Lorentz factor of radiating electrons to a value, which corresponds to the border between Thomson and {KN} 
Comptonization regimes for their own synchrotron radiation.}

The paper is organized as follows. In Sect.~\ref{the_model} we describe the model, which we use in our analysis.
The method we use to fit the data to the model is described in  Sect.~\ref{sec:SED.fitting}.
We discuss the SED fitting to the observations of   GRB~190114C in Sect.~ \ref{sec:best-fit-190114C}. 
In Sect.~\ref{sec:comparison} we compare our findings for GRB~190114C with some previous works. We then compare our findings with the predictions of  the pair balance model 
in Sect. ~\ref{sec:Pair.balance}. We 
summarize implications of our results in Sect.~\ref{sec:summary} and our conclusions in Sect.~\ref{sec:claims}.

\section{The model} 
\label{the_model}

\subsection{One-zone SSC {kinetic equations} }
One-zone model 
is the common simplest description of the  emitting region. It assumes that the region is uniform and isotropic. Thus, the distribution functions for all particle species depend only on their energy and, in non-stationary cases, on time. 
The radiating particles are injected into the zone {and} escape from it after producing photons on their way. Introduction of particle escape terms of the form $\left( {\partial N}/{\partial t} \right)_\mathrm{esc} = - {N}/{t_\mathrm{esc}}$ approximates this situation. Note that the effective escape times for radiating particles and photons {are, generally speaking,} different and this is accounted for by a geometrical factor $\Lambda$ {introduced} below.  If the   emitting region is expanding, then radiating particles undergo adiabatic cooling in addition to radiative losses.  This situation is approximated by adding expansion terms of the form $\left( {\partial N}/{\partial t} \right)_\mathrm{exp} = - {N}/{t_\mathrm{exp}}$ (where $t_\mathrm{exp} = V/\dot{V}$ and $V$ is the comoving volume) in combination with adiabatic cooling term for evolution of radiating particle's momentum, $\left(\mathrm{d} P/\mathrm{d} t\right)_\mathrm{ad} = - P/(3\,t_\mathrm{exp})$. 

We include the following  physical processes: 
\\ 
(i) Radiative synchrotron losses, that  are treated as continuous  emission and enter equations as the source term for photons and cooling term  for the electrons.
\\
(ii) Comptonization,  that can proceed in the  KN  regime. {Therefore, we} avoid continuous-loss approximation, {and} use instead  a collision integral for two-body collisions of electrons/positrons with photons with exact {QED} cross-section. Compton scattering is taken into account by two matching collision terms in equations for the photons 
{and}  radiating particles distributions.
\\
(iii) Two-photon pair production,  that is important when Comptonization proceeds in the  KN  regime and cooling due to IC losses is fast  compared to the effective photon escape time. Two-photon pair production is taken into account by two matching integral terms -- the drain term in equation for the photon distribution and a source term (which sums up with injection) in the equation for the distribution of radiating particles. Because of pair creation we have both electrons and positrons, and we group them together as ``radiating particles'' or simply electrons. {Radiation from secondary pairs} is included in our model and, as we show later, it plays an important role in fitting the spectrum of GRB~190114C. 

We neglect pair annihilation which is unimportant for GRB afterglows.  We also neglect synchrotron self absorption and induced scattering and therefore our calculations are not valid below the self absorption frequency, which is not relevant for the analysis carried out here. Finally we neglect collective plasma effects.

The  electrons' distribution functions, $f_\mathrm{e}(\gamma) = \mathrm{d} n_\mathrm{e}/\mathrm{d} \gamma$, where $\gamma$ is the electron Lorentz factor, satisfies, 
{within this formulation}:
\begin{equation} \label{electron_equation}
    \frac{\partial f_\mathrm{e}}{\partial t} = 
    - \frac{\partial}{\partial \gamma} \left( \dot{\gamma} f_\mathrm{e} \right) 
    + S_\mathrm{e} + Q_\mathrm{inj} + Q_\mathrm{pp}^\mathrm{(e)}  - \frac{f_\mathrm{e}}{t_\mathrm{eff}} \, ,
\end{equation}
where $t_\mathrm{eff}$ is the effective electrons' lifetime, $1/t_\mathrm{eff} = 1/t_\mathrm{esc} + 1/t_\mathrm{exp}$, $Q_\mathrm{inj}$ is the electrons injection rate, $Q_\mathrm{pp}^\mathrm{(e)}$ is the pair production rate, $S_\mathrm{e}$ describes Comptonization losses that are treated via electron-photon collision integral with exact QED cross-section.
$\dot \gamma$ is  the time derivative of $\gamma$ due to synchrotron losses and adiabatic cooling (note that IC cooling is treated separately in $S_e$): 
\begin{equation} \label{individual_losses}
    \dot{\gamma} = 
    - (\gamma^2 - 1) \frac{\sigma_{_\mathrm{T}}}{m_e c} \frac{B_\mathrm{rms}^2}{6\pi} 
    -  \frac{(\gamma^2 - 1)}{3 \gamma\, t_\mathrm{exp}}  \, .
\end{equation}
For the photons { distribution function}, $f_\mathrm{ph}(\epsilon) = \mathrm{d} n_\mathrm{ph}/\mathrm{d} \epsilon$,  where $\epsilon = E/(m_e c^2)$ is the dimensionless photon energy, we have:
\begin{equation} \label{photon_equation}
    \frac{\partial f_\mathrm{ph}}{\partial t} = Q_\mathrm{sy} +
    S_\mathrm{ph} - Q_\mathrm{pp}^\mathrm{(ph)}   - \frac{f_\mathrm{ph}}{\Lambda\, t_\mathrm{eff}} \, ,
\end{equation}
where $Q_\mathrm{sy}$ and $S_\mathrm{ph}$ are the synchrotron and IC photon sources.  The geometric factor $\Lambda$ reflects
the difference in effective lifetimes for photons and electrons. It arises mainly due to 
the inevitable anisotropy in the photon distribution within the source. In application to GRB afterglows (i.e., relativistic expanding blast wave) the geometric factor is $\Lambda \approx 1$ in the slow cooling regime and it increases logarithmically to $\sim$ several in the fast cooling regime \citep{PairBalance,DerishevPiran2019}. 

\subsection{Hydrodynamics and model coefficients } \label{sec:Coefficients}

The  kinetic description is  supplemented by a description of the hydrodynamic evolution of the  relativistic blast wave. The coupling between the hydrodynamics and the radiation is described by several dimensionless coefficients that are included implicitly in the formulation of the problem \citep{AfterglowModelling}. The one-zone nature of the model implies using average (in some sense) values for the shock's radius, kinetic energy of the shocked material,  the lifetime of emitting electrons, the Doppler factor of the emitted photons, and source-frame luminosity integrated over expanding spherical shock. 

In general, these quantities can be expressed in terms of  the shock front Lorentz factor $\Gamma$ and observer's time {$t_\mathrm{obs}$}.    
\cite{AfterglowModelling} express everything in terms of the Lorentz factor of the shocked material, denoted  $\gamma~(=\Gamma/\sqrt{2})$ in that paper
\citep[following][]{BlandfordMcKee}.
However,  some authors use  the \cite{AfterglowModelling} formulae, but with $\Gamma$  instead of $\gamma$ and this is a source of confusion.

We define the  factors $C_\mathrm{i}$ as:  
\begin{subequations}
\label{CoefficientDefinitions}
\begin{align} 
&    R = C_\mathrm{_R} \Gamma^2 c {t_\mathrm{obs} / (1+z)}  \label{emission_radius} \\
&    E_\mathrm{kin} = C_\mathrm{_E} \Gamma^2 M c^2 \\
&    t_\mathrm{{eff}} = {C_\mathrm{t} \Gamma {t_\mathrm{obs} / (1+z)}} \\   
&    h\nu_\mathrm{obs}  = C_{_\Gamma} \Gamma h\nu {/ (1+z)} \\
& { L = C_\mathrm{_L} \epsilon_r \; (1+z) E_\mathrm{kin}/t_\mathrm{obs}  }  \ , 
\end{align}
\end{subequations}
where $M$ is the swept-up mass and $h\nu_\mathrm{obs}$ the energy of observed photons, Lorentz boosted from $h\nu$ in the emitting zone comoving frame, {and $\epsilon_\mathrm{r}$ is the fraction of energy dissipated to  radiation}.
The coefficients depend on the hydrodynamics  \citep[e.g][]{BlandfordMcKee} that depends in turn on the density distribution of the circum-burst matter and on the radiative efficiency of the blast wave. But they also depend on different assumptions concerning averaging  within the emitting zone.

At times one needs to know the effective Lorentz factor of emitting matter, $\Gamma_\mathrm{em}$, i.e., the Lorentz factor that fluid immediately behind the shock front had at the moment when the {shock's radius was equal} to the effective emission radius (Eq.~\ref{emission_radius}), whereas the shock itself {had} 
already expanded to radius $R_\mathrm{sh} = 2(m+1) \Gamma^2 ct_\mathrm{obs} / (1+z)$, defined by self-similar \cite{BlandfordMcKee} solution, $\Gamma^2 \propto r^{-m}$,  where $m=1$ and $m=3$ for an adiabatic blast wave propagating into wind-like and ISM density profiles. {With the \new{} that we use in this paper, the} 
Lorentz factor of the emitting matter and the shock
front Lorentz factor are related as
\begin{equation}
\label{EmissionGamma} 
\Gamma_\mathrm{em} = \left( \frac{2(m+1)}{C_\mathrm{_R}} \right)^{m/2}  \frac{\Gamma}{\sqrt{2}}  \ .
\end{equation}

Numerous  authors considered different values for $C_\mathrm{_R}$. Specifically, for the ISM case:
\cite{Waxman97} gives $C_\mathrm{_R} = 1$,
\cite{Sari97} gives $C_\mathrm{_R} = 8$.
In the wind case:
\cite{DaiLu98} gives $C_\mathrm{_R} = 4$. 
\cite{PanaitescuMeszaros98b} calculated luminosity-averaged factors $C_\mathrm{_R}$ for both ISM and wind cases.
Other factors are usually hidden and not discussed. 
\cite{AfterglowModelling} give, either explicitly or implicitly, the full set of coefficients $C_\mathrm{i}$ for the ISM case.
Finally, \cite{newcode} calculated effective
values for all the coefficients for both ISM and wind cases. 
We use this last set of values (\new{}, see Table~\ref{CoefficientSet}) in our best fits. 
It is important to state, however, that the specific values of these coefficients change the inferred values of the physical parameters, but they don't change the quality of the fit  and qualitative characteristics of the solution.

Some choices of coefficients (\ref{CoefficientDefinitions}) are summarized in Table~\ref{CoefficientSet}. 
When attempting to reproduce fits obtained for GRB~190114C by other authors, we use the values  given by \cite{AfterglowModelling} {(\SPN{}, see Table~\ref{CoefficientSet})}.
{In the latter paper, as well as in that many  
others, the authors consider Eq.~(\ref{emission_radius}) as expression for both the effective emission radius and the shock's radius at the moment of observation. This implies a relation between shock's radius, its Lorentz factor and the upstream density, that does not comply with hydrodynamic solution of \cite{BlandfordMcKee} unless $C_\mathrm{_R} = 8$ (ISM case) or $C_\mathrm{_R} = 4$ (wind case). We follow this practice when trying to reproduce results of other authors, but our own best fit solutions are obtained taking into account the difference between the effective emission radius and the shock's radius. }

\begin{table*} 
\caption{Coefficients used in different one-zone afterglow  models. Note that in most cases these coefficients are introduced implicitly rather than explicitly in the corresponding papers. If a particular coefficient does not appear in a paper, neither explicitly nor implicitly, then it is not listed in the table.
}
\label{CoefficientSet}
\begin{tabular}{|p{0.3\textwidth}|p{0.23\textwidth}|p{0.34\textwidth}|}
\hline
   & \multicolumn{2}{|c|}{density profile}\\
\cline{2-3}\\[-5ex]
Reference & Wind & ISM \\
\hline
\parbox{0.3\textwidth}{\cite{AfterglowModelling},\\ ``\SPN{}'' \ hereafter} & 
&
$C_\mathrm{_R} = 2$,\,   $C_\mathrm{t} = 1/\sqrt{2}$,\,   $C_{_\Gamma} = 1/\sqrt{2}$,\,   {$C_\mathrm{_L} = 17/12\,^*$}
\\
\hline
\cite{PanaitescuMeszaros98b} & 
$C_\mathrm{_R} \approx 3.1$
&
$C_\mathrm{_R} \approx 6.5$ 
\\
\hline
\cite{Nava+13} & 
$C_\mathrm{_R} = 4/5$
&
$C_\mathrm{_R} = 8/9$ 
\\
\hline
\cite{DaiLu98} & 
$C_\mathrm{_R} = 4$,\,   $C_\mathrm{t} = 8\sqrt{2}/3$
&
$C_\mathrm{_R} = 8$,\,   $C_\mathrm{t} = 16\sqrt{2}/5$
\\
\hline
\cite{DerishevPiran2019} &
$C_\mathrm{_R} = 4$,\,   $C_\mathrm{_E} = 1$,\,   $C_\mathrm{t} = 4$,\,   $C_{_\Gamma} = 1$
&
$C_\mathrm{_R} = 8$,\,   $C_\mathrm{_E} = 1$,\,   $C_\mathrm{t} = 8$,\,   $C_{_\Gamma} = 1$
\\
\hline
\parbox{0.3\textwidth}{Current work \citep[from][]{newcode},\\ ``\new{}'' \ hereafter}
&
\parbox{0.35\textwidth}{$C_{_\mathrm{R}} \approx 2.45$,\,   $C_\mathrm{_E} = 2/9$,\,   $C_\mathrm{t} \approx 1.16$,\,   \\
$C_{_\Gamma} \approx 0.64$,\,  {$C_\mathrm{_L} = 9/8$}  }
&
\parbox{0.35\textwidth}{$C_{_\mathrm{R}} \approx 5.55$,\,   $C_\mathrm{_E} = 6/17$,\,   $C_\mathrm{t} \approx 0.96$,\,   \\
$C_{_\Gamma} \approx 0.87$,\,   {$C_\mathrm{_L} = 17/16$}  }
\\
\hline
\end{tabular}
\\

$^*$ This coefficient does not appear in the paper, but rather calculated using the same approach, that was used to normalize the distribution of emitting electrons.
\\
\end{table*}

Various authors, following \cite{GranotSari}  carried out ``beyond one-zone'' calculations in attempts to estimate the effects of a more realistic geometry on the resulting spectrum. However, as far as we know there was no systematic attempt to quantify these effects in terms of the above set of {coefficients} and explore their effect on the resulting spectrum.

\subsection{The injection function}
\label{sec:truncated_vs_smooth}

Following \cite{AfterglowModelling}, the common choice for the injection function  has been a power-law with a lower cutoff:
\begin{equation} \label{truncated_injection}
    Q_{\rm inj} (\gamma) \propto \left\{
    \begin{array}{ll}
        0 \, , &  \gamma < \gamma_\mathrm{m} \\
        \gamma^{-p} \, , &  \gamma \geq \gamma_\mathrm{m}  ,
    \end{array}
    \right.
      \quad\gamma_\mathrm{m} = \bigg(\frac{p-2}{p-1}\bigg) \, \langle \gamma \rangle
    \; ,
\end{equation}
where $\langle \gamma \rangle$
is the average Lorentz factor of the injected electrons.

This truncated injection function was introduced in the context of analytic approximations 
and it allows  to reduce the calculations to finding simple asymptotic forms while keeping all the essential physics.
Numerical simulations allow more complicated expressions and it makes sense to use a smooth 
form of the injection function. {Therefore, in this paper we use}
the form {introduced in \citep{PairBalance}, }
\begin{align} \label{smooth_injection}
    &Q_{\rm inj} (\gamma) \propto \frac{{\gamma^2-1}}{\left( \gamma_{\rm b} + \gamma \right)^{p+2}} \ ,\\
    &{\rm where ~~}  
     \gamma_{\rm b} = \frac{p-2}{3}  \langle \gamma \rangle \; {\rm ~~~(for ~~} \gamma_{\rm b} \gg 1). \nonumber
\end{align}
This function is differentiable in momentum space for all $\gamma$ and converges to the expression (\ref{truncated_injection}) in the limit $\gamma \gg \gamma_\mathrm{b}$. Being smooth, it allows one to avoid numerical artefacts, which originate from the step discontinuity in truncated injection function.

Usually the average Lorentz factor of the injected electrons is expressed in terms of the energy fraction in accelerated electrons,
\begin{equation} \label{avg_gamma}
    \langle \gamma \rangle
    =  \frac{\epse}{\xi_e}  \frac{\mu}{m_e}  \frac{\Gamma}{\sqrt{2}} \ , 
\end{equation}
where $\mu$ the mass per one {upstream} electron ($\mu = m_p$ for purely hydrogen composition). 
{The fraction of upstream electrons that are being injected, $\xi_e$, is commonly assumed to be unity. However, in the pair-balance model the injected particles originate from  secondary electron-positron pairs created in the upstream, so one generally expects $\xi_e \ne 1$ (even $\xi_e > 1$ is possible).}

Whatever is the injection function, we supplement it by exponential cut-off at the Lorentz factor
\begin{equation}
\gamma_{\rm max} \approx   \left( \frac{6 \pi q_e}{\sigma_T B} \right)^{1/2} \ , 
\label{eq:gamma_max}
\end{equation}
such that the synchrotron cooling rate equals the maximum acceleration rate $q_e B c$ and further acceleration is problematic. 
Here
 $q_e $ is the electron's charge, $\sigma_T$ is the Thompson cross section and $B$ is the downstream magnetic field. 
IC losses may decrease $\gamma_{\rm max}$. However, it happens only if IC losses are dominant even for highest-energy electrons in spite of possibly large  KN  corrections. This is not the case for GRB afterglows.

\section{SED Fitting}
\label{sec:SED.fitting}

We explore a stationary one-zone SSC model with effective (i.e., average) parameters to approximate  the observed spectrum of a decelerating blast wave. 

To run ahead of ourselves a bit,  this leads, for GRB~190114C, to a surprisingly good SED fit from the visible band to TeV. 
We estimate the goodness of SED fit in the standard way, by calculating $\chi^2$ with the set of XRT, BAT, LAT, and MAGIC observational data points; all of them are taken from \cite{MAGIC_Nature_fit}.
Note  that the observed spectra are not instantaneous but collected over a finite time interval. 

\cite{MAGIC_Nature_fit} also provide a table of Swift-UVOT measurements taken with white filter and an estimate for the host galaxy extinction ($A_{_V} = 1.83 \pm 0.15$). With these measurements we calculated the average magnitudes in two time intervals, $68 \div 110$~s and $110 \div 180$~s, which are $14.00 \pm 0.03$ and $14.91 \pm 0.03$ respectively. Then we converted these magnitudes into $\nu F_{\nu}$ values at 2~eV using UVOT white filter response\footnote{\url{http://svo2.cab.inta-csic.es/svo/theory/fps3/index.php?id=Swift/UVOT.white}}, correcting for extinction in our galaxy 
\cite[$A_{_V} = 0.035,$][]{GalacticExtinction}, 
applying \cite{ExtinctionModel} model for host galaxy extinction, and assuming spectral index $\beta = -0.2$ ($F_{\nu} \propto \nu^{\beta}$), which agrees
with \cite{MAGIC_Nature_fit} estimate, $\beta = -0.10 \pm 0.12$. This procedure resulted in average $\nu F_{\nu} (2\ \mathrm{eV}) \simeq 1.14 \times 10^{-9}$~erg/s/cm$^2$ for $68 \div 110$~s time interval and $\nu F_{\nu} (2\ \mathrm{eV}) \simeq 4.9 \times 10^{-9}$~erg/s/cm$^2$ for $110 \div 180$~s time interval.
Systematic errors due to uncertainty in the host galaxy extinction are $-25\% +30\%$.

The one-zone SSC model is described by 4 macroscopic and 2 microscopic parameters. The macroscopic parameters are the shock's Lorentz factor $\Gamma$, the observation time $t_\mathrm{obs}$,  
which in turn together with $\Gamma$ determines the life time of particles in the emitting zone (see Section  \ref{sec:Coefficients}), the
magnetic field strength, and the energy density of the accelerated (injected) electrons/positrons. The last two quantities are typically expressed in terms of the equipartition parameters $\epsB$ and $\epse$ in combination with the energy density of the shocked material, which in turn is proportional to the isotropic equivalent kinetic energy of the shock, $E_{\rm kin}$. 
The shock's kinetic energy merely sets the typical GRB energy scale and has no impact on the calculated SEDs. Its value is  arbitrary to a certain extent, although there are lower and upper limits based on  
the total observed flux and the physical model of the radiating relativistic shock. 
The two microscopic parameters determine the energy distribution of the leptons, that we characterize by a lower-energy cutoff at $\gamma_{\rm b}$ (or $\gamma_\mathrm{m}$) and a high energy spectral index of the distribution, $p$.

This characterization of the problem depends on the blast wave only in two ways. First, as mentioned earlier, $t_{\rm obs}$ (and $\Gamma$) determine the leptons cooling time. Second, we take into account averaging over the emitting region that is reflected in the coefficients discussed  in section \ref{sec:Coefficients}. 
We use the set of  \new{}. In section (\ref{sec:model-variations}) we compare to the results obtained with the more familiar \SPN{}. Using different coefficients leads to virtually the same SED fit but of course the implied parameters will be somewhat different.

Our approach is to perform a systematic scan over the 4-dimensional parameter space (the  free parameters are $\Gamma$, $B$, $\gamma_{\rm b}$ (or $\gamma_\mathrm{m}$), and the ratio of leptonic to magnetic energy $\Ksc$)
looking for all regions where the calculated SEDs fit the observations. We do not consider the power-law index of the injection function as a free parameter and expect that its value is universal and is dictated by the particle acceleration process.  When calculating the SSC spectra, we use the value $p=2.5$ predicted in the pair balance model \citep{PairBalance}. However, in section (\ref{sec:injection-index}) we discuss the sensitivity of our conclusions to the particular choice of $p$. 
At the same time we differ from the traditional approach by treating $\gamma_{\rm b}$ as independent free parameter instead of choosing the value proportional to $\Gamma$ \citep{AfterglowModelling}. This is motivated by the prediction of the pair-balance model \citep{PairBalance}, where $\gamma_{\rm b} \propto B^{1/3}$.  To allow for these (and maybe other) possibilities we relax the model-dependent assumption  $\gamma_{\rm b} \propto \Gamma$. 
We use smooth injection {(Eq.~\ref{smooth_injection}),} but in {Sect.~\ref{sec:SPN-truncated}} we compare to the results obtained with  truncated injection {(Eq.~\ref{truncated_injection}).} 

We calculate the SSC spectra numerically, using a newly-developed single-zone code \citep{newcode}, which evaluates both the underlying electron/positron distribution and the resulting photon distribution.
The code takes into account synchrotron emission considering inhomogeneity of the magnetic field (assuming Gaussian distribution for its local strength), inverse Compton scattering with recoil using the  exact QED cross-section, and two-photon production of electron-positron pairs (re-injected into the emitting zone), also with the exact cross-section.  For a given stationary injection the code finds the instantaneous SED.

When scanning over the parameter space  we fix two out of four parameters used in the modelling and then vary the other two until we find the SED with minimal $\chi^2$ achievable for the pair of fixed parameters. This procedure is straightforward when one varies $\gamma_{\rm b}$ and $\Ksc$ with fixed $\Gamma$ and $B$ --- there is only one minimum in the function $\chi^2(\gamma_{\rm b},\Ksc,B=const,\Gamma=const)$. The functions  $\chi^2(\gamma_{\rm b},B,\Ksc=const,\Gamma=const)$ and  $\chi^2(\Ksc,B,\gamma_{\rm b}=const,\Gamma=const)$ have, generally speaking, several local minima. However, in the case of GRB~190114C one of the minima is well below the others, and we consider only this one.

\section{Modelling GRB~190114C}
\label{sec:best-fit-190114C}
We turn now to apply our method to fit the observed spectrum of GRB~190114C.
The results of parameter space scans are presented in the following subsections. The case of the early-time (68 to 110 s after the trigger) SED fit is described in greater detail. It serves both as a practical example of our approach and as the reference for further analysis. {We find the best fit parameters for the early time and the best fit parameters for the late time as well as joint fit parameters corresponding to an adiabatic evolution between the early and late time. While the parameters groups are not that different from each other, we use the latter as our ``preferred set of parameters" (see Table~\ref{ParametersSummary}). }

\subsection{The  best-fit SED at early time}
\label{sec:pivotal-early-time}

We approximate the earliest available data set, 68 to 110~s after the trigger, with instantaneous SED calculated for $t_{\rm obs} = 90$~s. Even though we fit an instantaneous spectrum, 
the \new{} used in the fit  depend on the blast wave history. This in turn depends on the {circumburst}  density profile.  The results presented in this sub-section are for a blast wave propagating {into} a wind external density {profile}. The ISM case is discussed in Sec.~\ref{sec:model-variations}. When considering instantaneous spectra, 
{the different coefficients  lead to slightly different inferred physical parameters, but the SED itself remains essentially the same.}

\begin{figure}
    \centering
 \includegraphics[width=0.95\columnwidth]{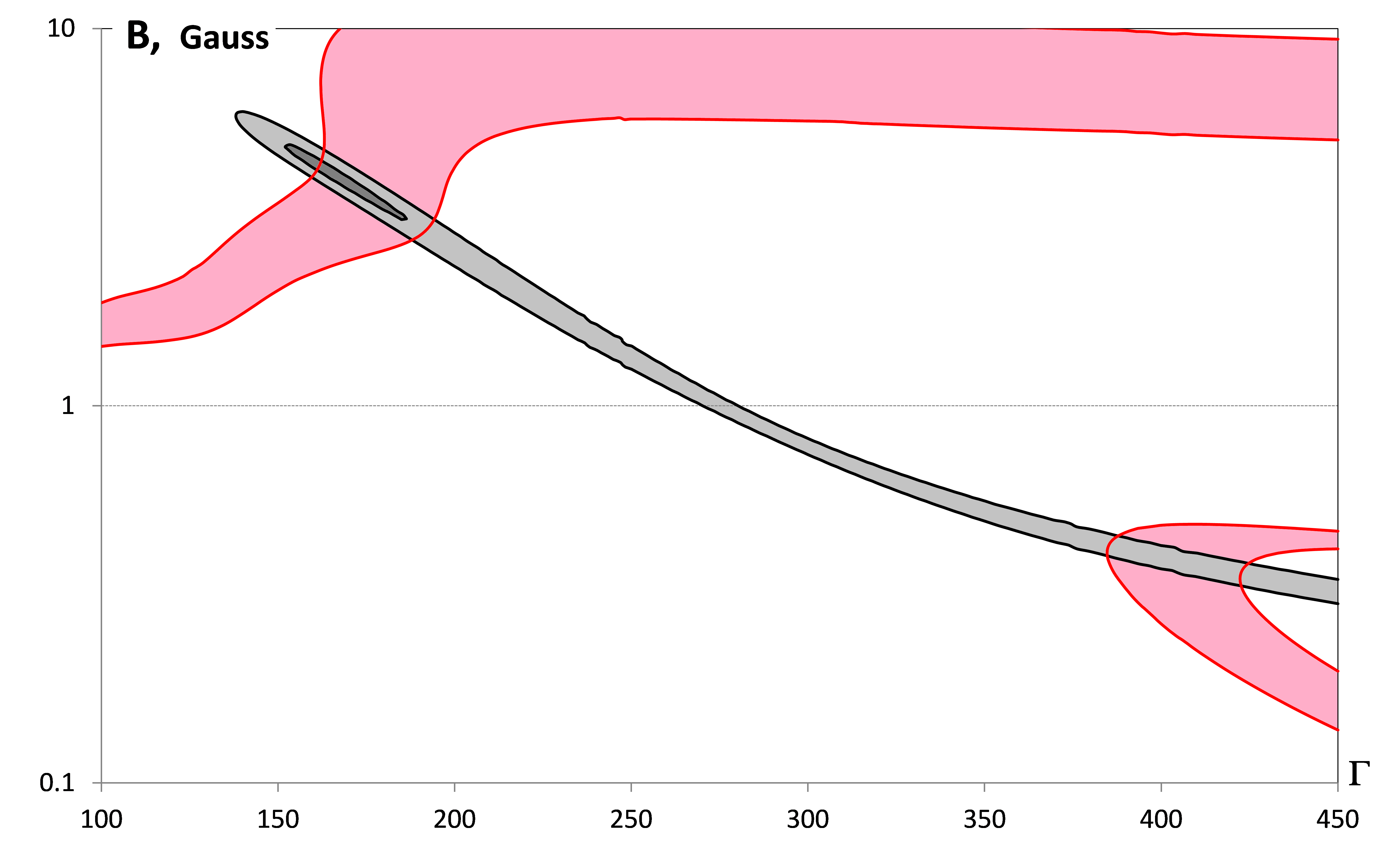}
  \includegraphics[width=0.95\columnwidth]{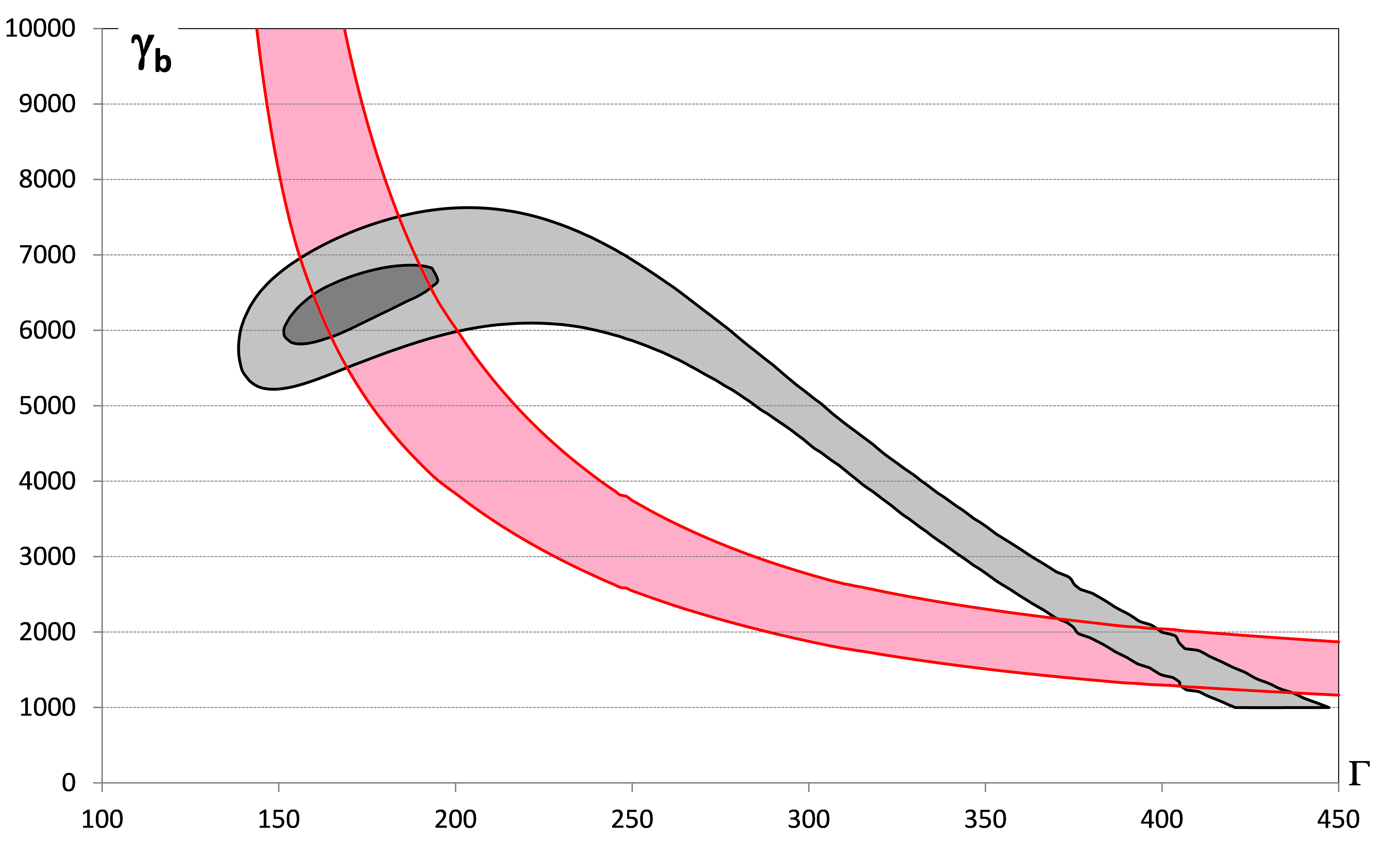}  
  \includegraphics[width=0.95\columnwidth]{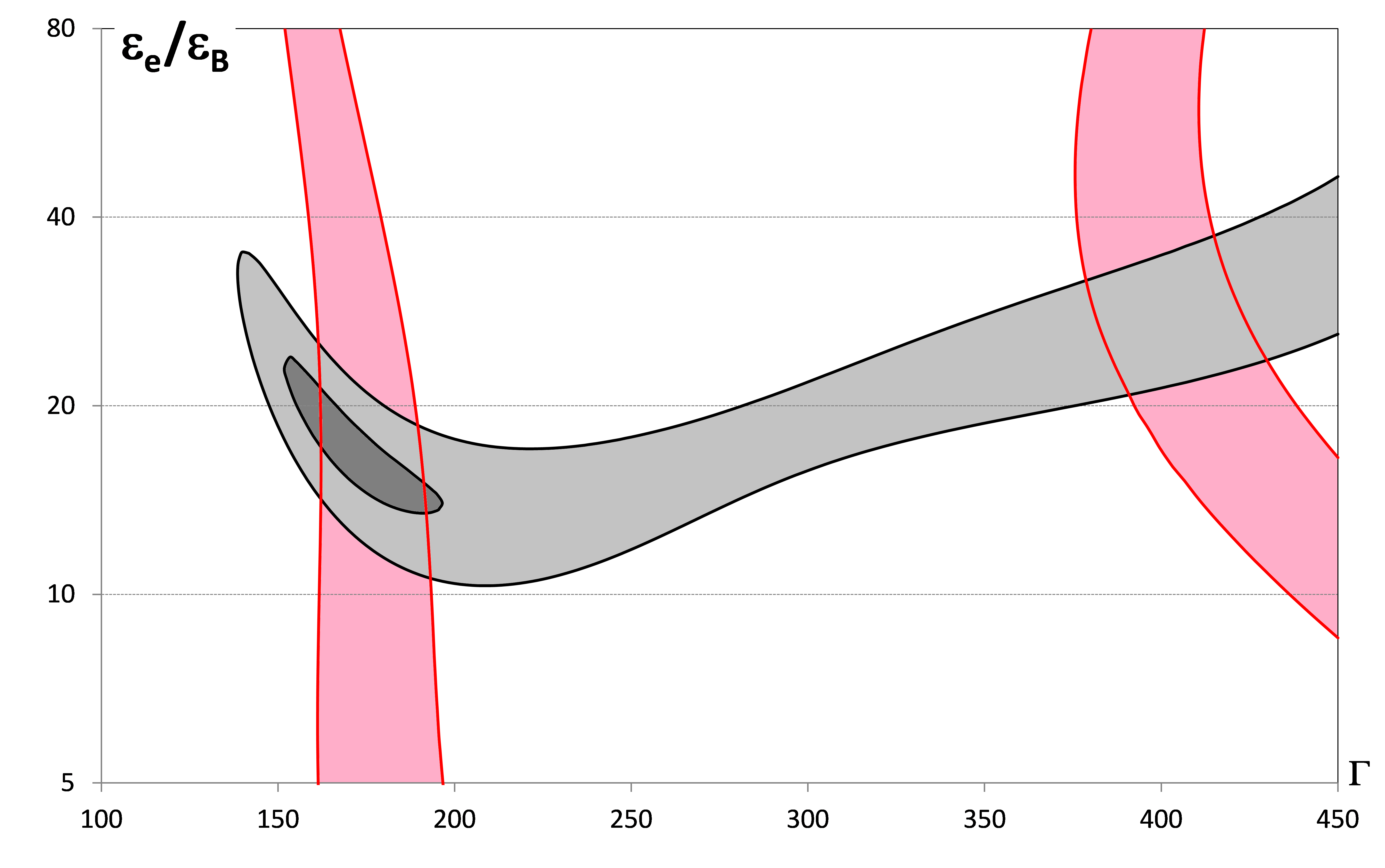} 
  \caption{Regions of good fits {projected onto 2D sections} of the {parameter} space. For each {pair} of parameters (e.g. {$\Gamma$ -- $B$} in the top panel)  we show {a projection of the surface that} passes through the {best-fit} values for the two other parameters ({i.e., $\gamma_\mathrm{b}$ and $\Ksc$ in the top} panel).  Shown are the regions of good   ($\chi^2 < 12$, dark grey)  and  acceptable  ($\chi^2 < 16$, light grey) fits to the TeV, GeV  and X-ray observations. The region where the calculated optical flux is {$0.7 \div 1.35$ of the observed value (covering both systematic and statistical uncertainties)} is colored light red. 
  {\bf Top}:   $B$ -- $\Gamma$     parameter plane. 
  {\bf Middle:}  $\gamma_\mathrm{b}$ -- $\Gamma$ parameter plane.
  {\bf    Bottom:} $\Ksc$ -- $\Gamma$ parameter plane.
  }
      \label{Fig.early_parameters_space}
\end{figure}

The best (minimal) SED $\chi^2$ values achieved in our parameter space scans are shown in {Fig.~\ref{Fig.early_parameters_space}.} Each point in the plots represents a solution obtained for two given parameters (say $\Gamma$ and $B$ {in the top panel}), where we optimize the fit over the two other variables ($\gamma_{\rm b}$ and $\Ksc$ in that case).
Three conclusions immediately follow from these plots.

First, the parameter space region, where we find reasonably good SED fits, {is} strongly elongated. {It extends} from $\Gamma \simeq 140$, that corresponds to the fast cooling regime, {upwards. There,} starting from $\Gamma \simeq 270$,  the solution becomes slow cooling. 
Although the best fit is firmly located at $\Gamma \simeq 165$, our fits can accommodate any $\Gamma \gtrsim 140$ within the range that we have explored ($100 \leq \Gamma \leq 450$), even though we have combined
X-ray, GeV and TeV observations.
On the lower $\Gamma$ values side, the allowed region is limited by internal two-photon opacity that, if becomes too high, eliminates (through electromagnetic cascade) the dip between synchrotron and IC peaks in SED, whose presence is evident from the LAT data and from the curvature of the X-ray spectrum. 
On the  high-$\Gamma$ side, the limiting factor is the hardness of the TeV tail, which, however has rather weak statistical support. 

Second, the most stringent lower limit on the shock's Lorentz factor, $\Gamma \gtrsim 160$, arises from the visible band observations.  These observations put 
 an  upper bound on the contribution of secondary pairs to  visible light. And in turn, this limits the amount of internal two-photon absorption and secondary pair creation  to a level much smaller than the one allowed by the IC peak shape.
Surprisingly, the most stringent upper limit on the shock's Lorentz factor, $\Gamma \lesssim 420$, also derives from the visible band observations. The large-$\Gamma$ fits are slow-cooling solutions and the position of X-ray peak is associated with the cooling Lorentz factor $\gamma_\mathrm{c}$, whereas $\gamma_\mathrm{b}$ has to be much smaller (as dictated by the location of IC peak). This eventually increases the power of the optical-band  emission to unacceptable high levels. 

Third, the parameters of the SED fit, that reproduces the observed visible light  flux in addition to X-ray and TeV spectra,  are close to those of the best fit based on X-ray to TeV data points alone. {Thus}, there exists a good SED fit that reproduces observations in a very broad range, from $\sim 1$~eV to $\sim 10^{12}$~eV. 
With this solution there is no need to invoke any additional source of  visible light photons, like an extra population of ``thermal'' electrons or a reverse shock.

A narrow valley of good fits extends from the global $\chi^2$ minimum near the lowest allowed shock's Lorentz factor towards the largest possible $\Gamma$. 
At both ends we find a SED fit that reproduces the observed visible band luminosity in addition to fitting the X-ray, GeV and TeV data points. The statistically significant global minimum of $\chi^2$ (with respect to the X-ray to TeV data) falls in the region with a good fit to the optical.  Solutions with lower and higher $\Gamma$ are ruled out by producing too much optical radiation. Any solution from the valley of good fits between the two optically consistent SEDs provides a reasonably good fit from X-rays to TeV, that is likely the reason why different attempts to fit SED of GRB~190114C in the literature resulted in different parameters. In the middle of the valley the predicted optical-band flux is so low that one needs another independent source of visible light.

\begin{figure}
    \centering
 \includegraphics[width=1.0\columnwidth]{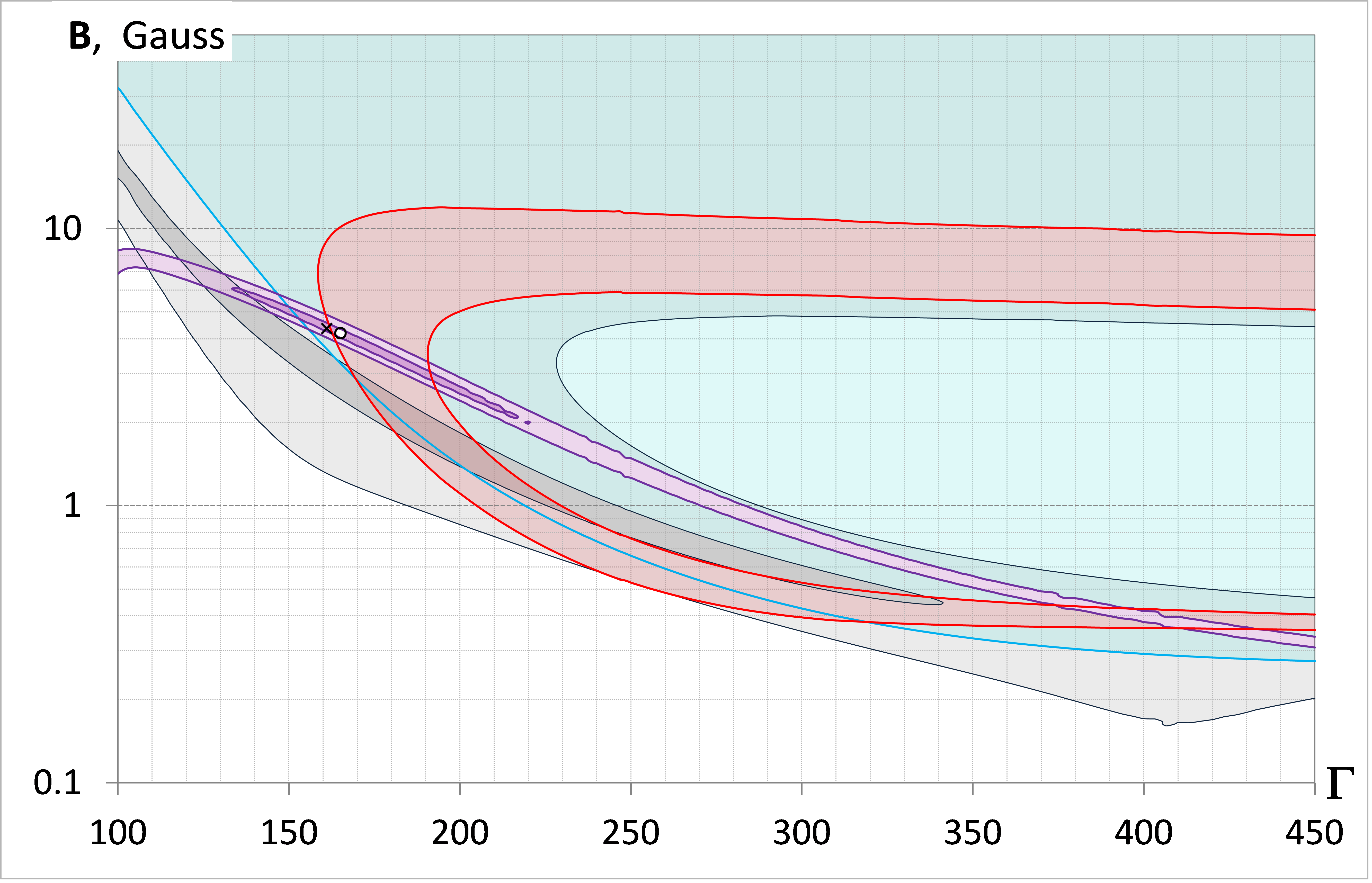}
    \caption{
    {The parameter plane in $B$--$\Gamma$ coordinates for $t_\mathrm{obs} = 90$~s, wind density profile, and smooth injection with $p=2.5$.}
    Regions  where good fits to  12  X-ray data points can be obtained by varying $\gamma_\mathrm{b}$ and $\Ksc$ {are colored gray} --- darker {shade} corresponds to $\chi^2_\mathrm{x-ray} < 3.4$, lighter {shade} corresponds to $\chi^2_\mathrm{x-ray} < 6$, and the absolute minimum is $\chi^2_\mathrm{x-ray} \simeq 3.2$.
    The blue shaded region corresponds to good fit to the singe GeV data point ($\chi^2_\mathrm{_{GeV}} < 1$). Magenta shades indicate goodness of fit for 6  sub-TeV data points  -- darker shade for $\chi^2_\mathrm{_{TeV}} < 8$ and lighter shade for $\chi^2_\mathrm{_{TeV}} < 12$.  The region where calculated optical flux is {$0.7 \div 1.35$ of the observed value (covering both systematic and statistical uncertainties)} is shaded light red.
    {The open circle shows the location of the best fit to X-ray, GeV and sub-TeV data points, and the nearby cross shows the solution that we choose taking into account later-time ($t_\mathrm{obs} = 145$~s) data points.}
    }
      \label{Xray_fit_plus_other_bands}
\end{figure}

It is interesting to analyze how restrictive is the requirement to fit the X-ray data points ignoring other information and how much observations in various spectral bands contribute to our ability to pin-point the GRB parameters. Figure~\ref{Xray_fit_plus_other_bands} demonstrates that the X-ray data are not at all restrictive -- we were able to find excellent SED fits for virtually any combination of shock's Lorentz factor and the magnetic field strength by varying just two remaining parameters, $\gamma_\mathrm{b}$ and $\Ksc$. 
When the  {(single)} GeV observation is added to the X-rays it  excludes some regions at outskirts, most notably at small $\Gamma$, but still  a huge degeneracy remains.
Optical measurement provides even stronger constraint on the lowest possible $\Gamma$ and excludes extremely strong magnetic field
(a similar observation was made for the prompt phase by \cite{BeniaminiPiran2014}).
Note that while a single-zone SED fit in the region inside the red arc 
does not reproduce the observed optical flux, it is still allowed if one considers an additional source of visible light. Eventually, the most stringent constraint comes from TeV-range data points -- the only observations that directly capture the IC component in the GRB spectra. However, degeneracy with respect to choosing the shock Lorentz factor persists unless one requires that the  model  reproduces the observed optical flux as well.  The intersection of TeV-allowed region with the optical band region is the critical factor that determines the parameters.

\begin{figure}
    \centering
 \includegraphics[width=1.0\columnwidth]{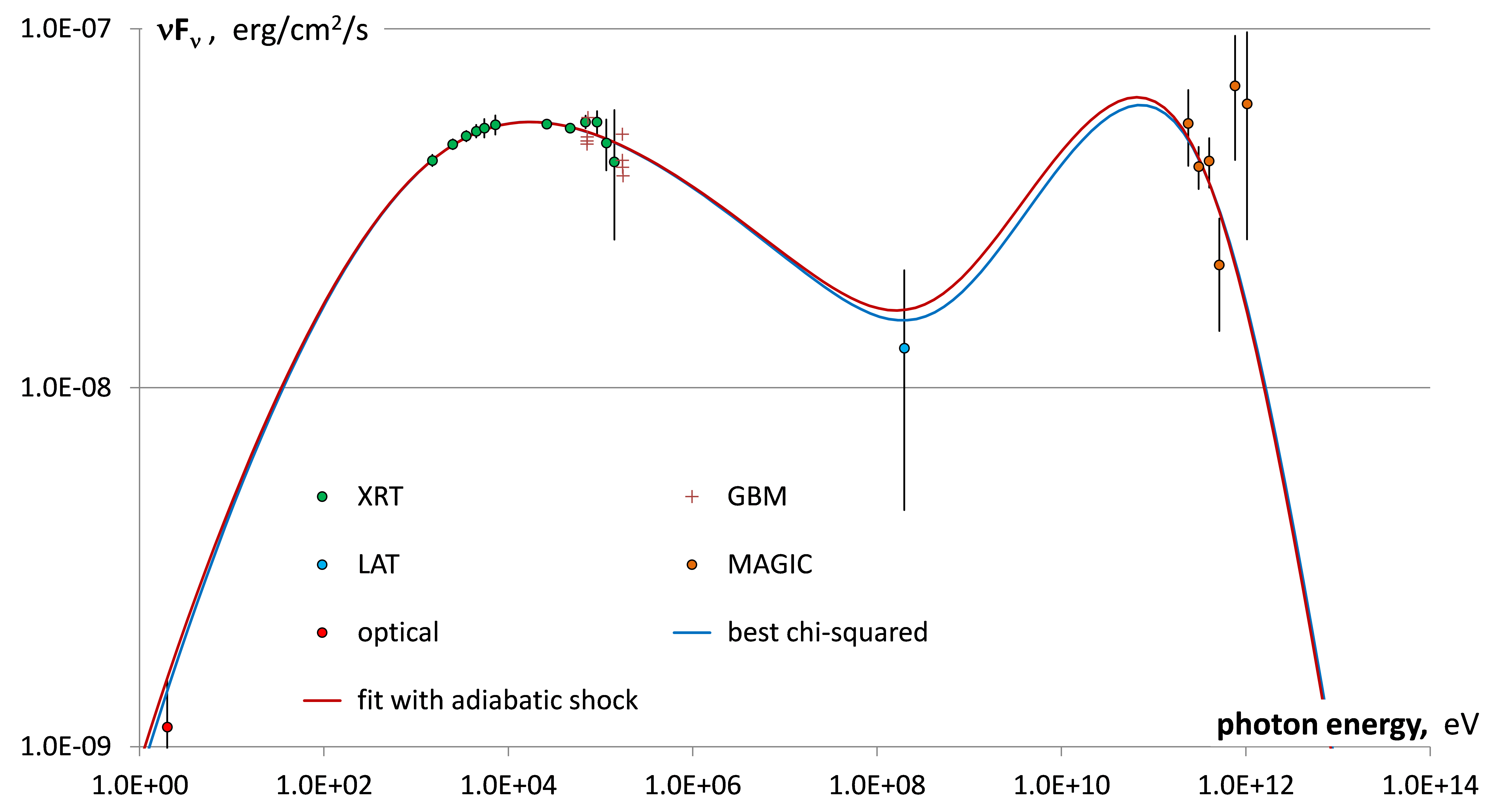}
 \includegraphics[width=1.0\columnwidth]{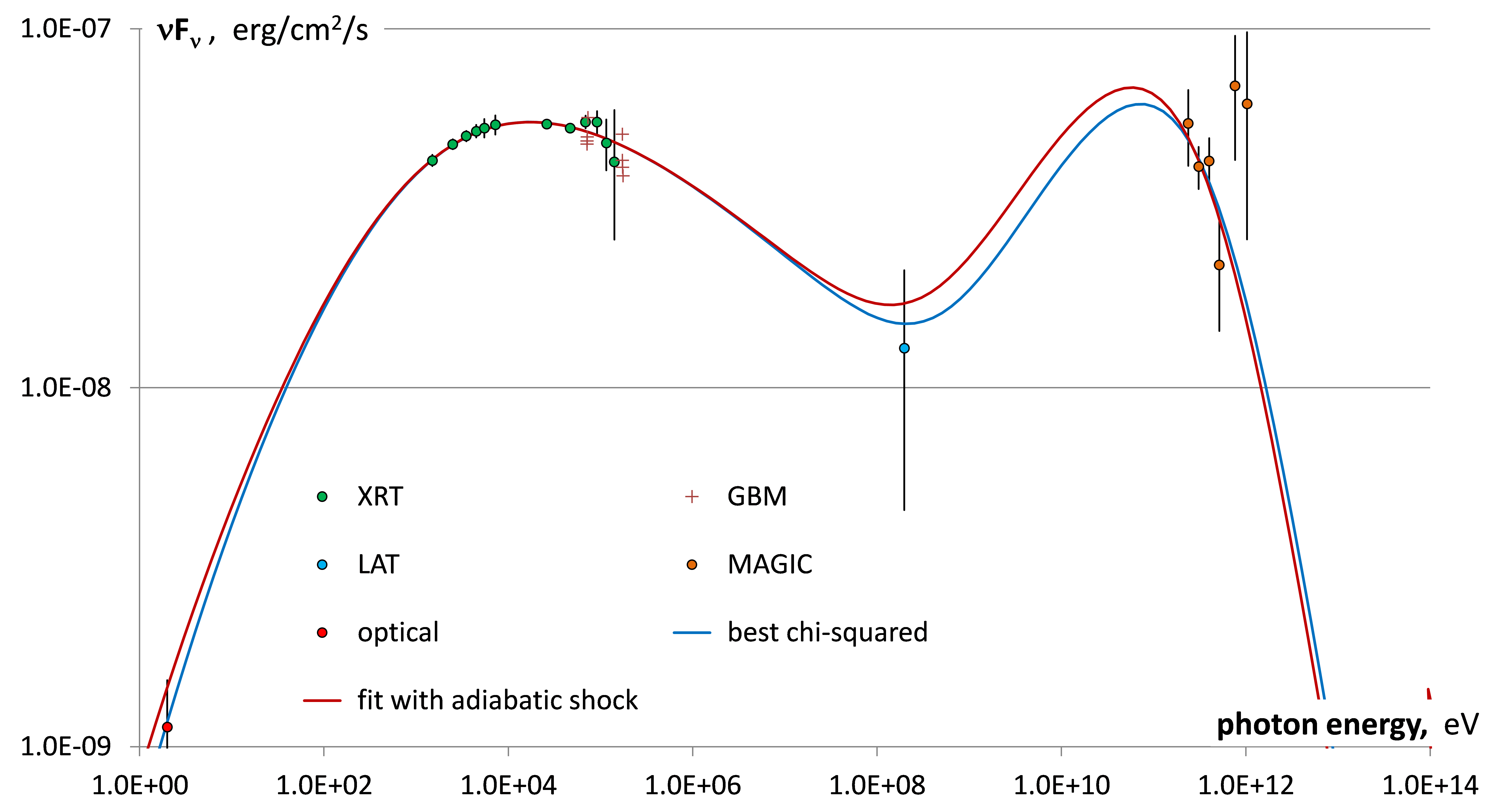}
    \caption{ 
    A comparison of the best $\chi^2$ SED fits to X-ray, GeV and TeV data points at $t_\mathrm{obs} = 90$~s (blue curves) to 
    SED fits that correspond to {adiabatic blast wave with} the same circumburst density parameters for both  early and  late observations (red curves) for progenitor's mass loss rate ($1.4 \times 10^{-6} V_\mathrm{w,3000} E_\mathrm{53.5}\  M_{\sun}$/yr) or fixed ISM density ($2\, E_\mathrm{53.5}\ m_p$/cm$^3$).
    {\bf Top}:  wind density profile. 
    Best fit parameters are $\Gamma = 165$, $B = 4.18$~G {($\epsB \simeq 6.4 \times 10^{-3} / E_\mathrm{53.5}$),} 
    $\Ksc \simeq 18.5$, $\gamma_\mathrm{b} \simeq 6620$, resulting in $\chi^2 \simeq 11.4$ for 19 {data points} (not counting the optical point) with four parameters.
    Fixed progenitor's mass-loss rate fit parameters are $\Gamma = 161$, $B = 4.37$~G {($\epsB \simeq 6.1 \times 10^{-3} / E_\mathrm{53.5}$),} 
    $\Ksc \simeq 19.6$, $\gamma_\mathrm{b} \simeq 6540$, resulting in $\chi^2 \simeq 11.4$ for 19 {data points} (not counting the optical point) with four parameters.
    {\bf Bottom}:   ISM density profile. 
    Best fit parameters are $\Gamma = 115.5$,  $B = 4.99$~G {($\epsB \simeq 6.7 \times 10^{-3} / E_\mathrm{53.5}$),} 
    $\Ksc \simeq 19.0$, $\gamma_\mathrm{b} \simeq 5570$, resulting in $\chi^2 \simeq 11.2$ for 19 {data points} (not counting the optical point) with four parameters.
    Fixed ISM density fit parameters are $\Gamma = 109$, $B = 5.68$~G {($\epsB \simeq 6.2 \times 10^{-3} / E_\mathrm{53.5}$),} 
    $\Ksc \simeq 21.4$, $\gamma_\mathrm{b} \simeq 5700$, resulting in $\chi^2 \simeq 11.5$ for 19 {data points} (not counting the optical point) with four parameters.
    }
      \label{fits_at_early_time}
\end{figure}

The best fit spectrum over the whole parameter space is shown  Fig.~\ref{fits_at_early_time}. 
It is possible to fit well the optical band flux at a price of a negligible increase of $\chi^2$ in X-ray to TeV range, however  keeping  in mind the necessity to fit the SED at a later  time with the same  circumburst medium parameters,  we choose a solution corresponding to smaller shock's Lorentz factor (and with slightly larger value of $\chi^2$) as our reference early-time SED fit.
In both  these fits $\gamma_\mathrm{b} \simeq \left( B_\mathrm{cr} /B \right)^{1/3}$, where $B_\mathrm{cr}\approx 4.4 \times 10^{13}$ Gauss is the critical (Schwinger) magnetic field.  This approximate equality is a coincidence within the common model, but is a prediction of the pair balance model.

The second  optically-consistent solution that we find at the  {high-$\Gamma$} end of the valley of good fits 
corresponds to the slow cooling regime. This solution
has markedly worse goodness of fit as compared to the lower-$\Gamma$ solution. It requires larger $\epsB$ 
(and, correspondingly, larger $\epse$). The very large $\Gamma$ in this solution implies unreasonably small circumburst density.

\subsection{Varying the model coefficients}
\label{sec:model-variations}

\subsubsection{ISM external density profile }
\begin{figure}
    \centering
 \includegraphics[width=1.0\columnwidth]{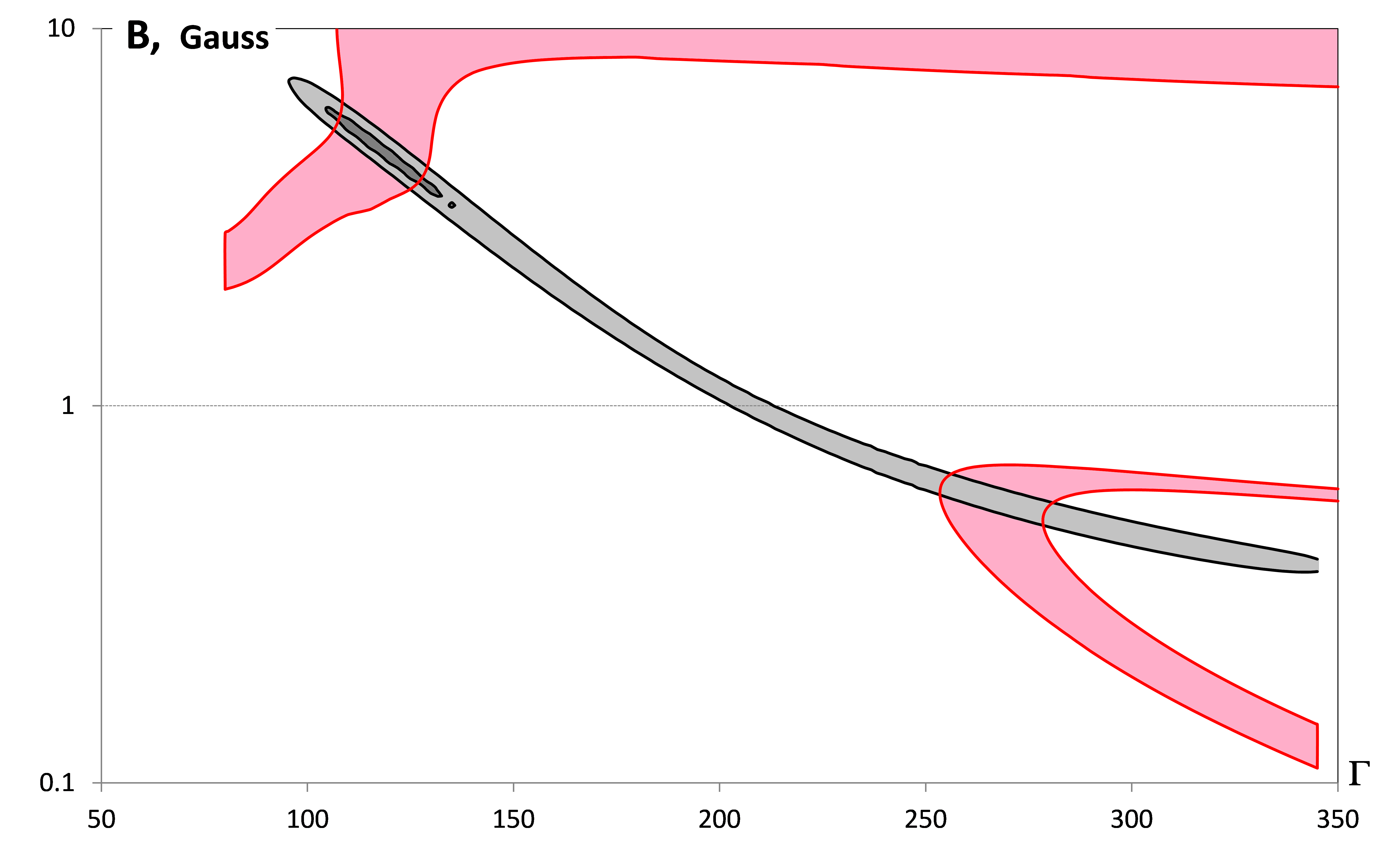}
    \caption{{Regions of good fits in $B$ -- $\Gamma$ parameter plane obtained using the \new{} for ISM density profile (i.e., same as the top panel of Fig.~\ref{Fig.early_parameters_space}, but for constant external density).  Shown are the regions of good   ($\chi^2 < 12$, dark grey)  and  acceptable  ($\chi^2 < 16$, light grey) fits to the TeV, GeV  and X-ray observations. The region where the calculated optical flux is $0.7 \div 1.35$ of the observed value (covering both systematic and statistical uncertainties) is colored light red.}}
      \label{ISM_parameter_map}
\end{figure}

Extending the results obtained in Sec.~\ref{sec:pivotal-early-time} for the wind-type external density profile to the case of constant external density (ISM case for short) requires nothing more than replacing the set of coefficients according to   Table~\ref{CoefficientSet}. Using the coefficients for ISM case we arrive at very similar results (see Fig.~\ref{ISM_parameter_map}): the allowed (good $\chi^2$ for X-ray, GeV and TeV data points) solutions are located within a narrow band extending from shock's Lorentz factor 
$\Gamma \simeq 95$ upwards, with fast-cooling (i.e., having {low} $\Gamma$) solutions being preferred\footnote{Like in the wind case high $\Gamma$ solutions are in the slow cooling regime.}, the best fit based on combination of X-ray, GeV and TeV data points {(see bottom panel of Fig.~\ref{fits_at_early_time})} is very close in parameter space to the solution that reproduces the observed optical flux.
{The} physical conditions in the emitting zone are similar to those obtained in the wind case. 
Note that  with the \new{} that we use, 
the effective emission radius is smaller than the shock's radius, and the Lorentz factor of the emitting matter at the effective emission radius is larger than {implied by} the shock front Lorentz factor (due to deceleration, {see Eq.~\ref{EmissionGamma}}); it is $\Gamma_\mathrm{em} \simeq 149$ in the wind case and $\Gamma_\mathrm{em} \simeq 141$ in the ISM case.

\subsubsection{Other coefficients}
\label{sec:SPN-truncated}
\begin{figure}
    \centering
 \includegraphics[width=1.0\columnwidth]{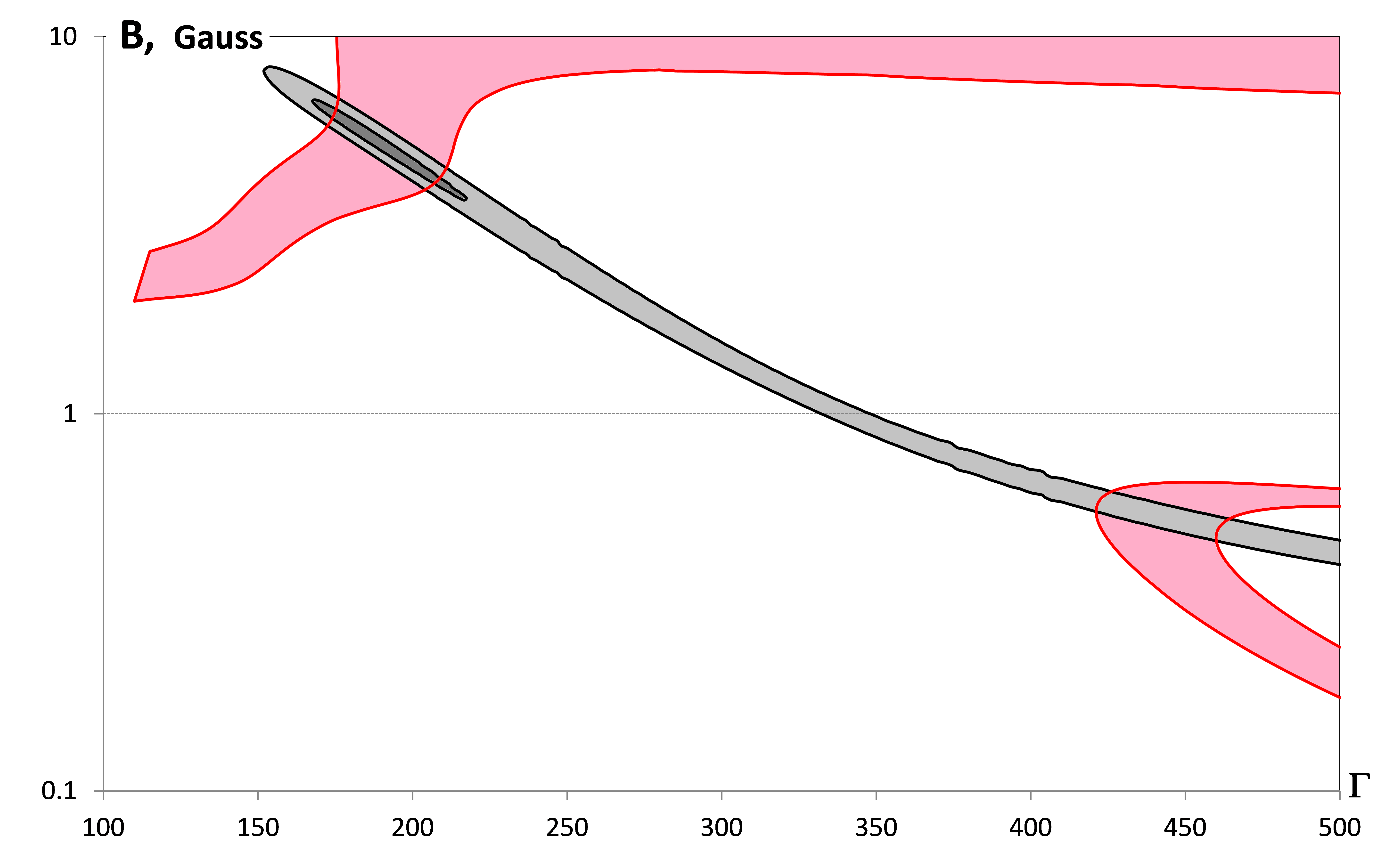}
    \caption{{Regions of good fits in $B$ -- $\Gamma$ parameter plane obtained using the \SPN{} for ISM density profile (i.e, same as  Fig.~\ref{ISM_parameter_map}, but for different set of coefficients).  Shown are the regions of good   ($\chi^2 < 12$, dark grey)  and  acceptable  ($\chi^2 < 16$, light grey) fits to the TeV, GeV  and X-ray observations. The region where the calculated optical flux is $0.7 \div 1.35$ of the observed value (covering both systematic and statistical uncertainties) is colored light red.}
    }
      \label{SPN_parameter_map}
\end{figure}

We also look for a best-fit solution using the common \SPN{} that correspond to the ISM case. 
The results, presented in Fig.~\ref{SPN_parameter_map}, are remarkably similar,  in terms of the topology of the allowed and preferred regions in the parameter space,  to the results obtained with the {\new{}.}  
The implied physical conditions in the emitting zone are different, but given the approximate nature of the one-zone model we do not consider this difference as significant.

\subsection{Different electrons' injection function}
\label{sec:injection-index}

\begin{figure}
    \centering
 \includegraphics[width=1.0\columnwidth]{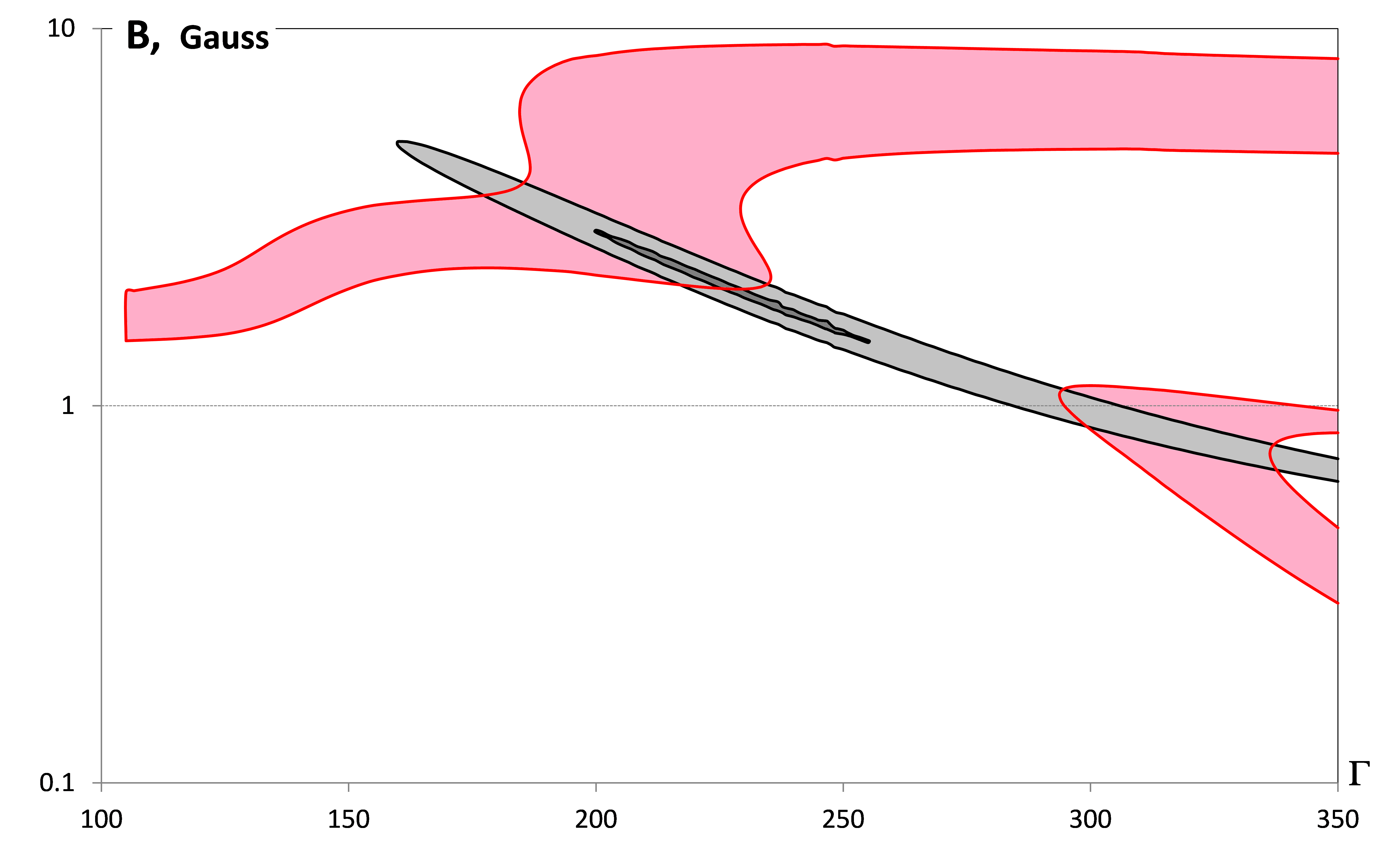}
    \caption{Regions of good fits ($\chi^2 < 12$, dark grey) and regions of acceptable fits ($\chi^2 < 16$, light grey) in $B$ -- $\Gamma$ parameter plane for hard injection with $p=2.3$. The region where calculated optical flux is $0.7 \div 1.35$ of the observed value (covering both systematic and statistical uncertainties) is colored light red.}
      \label{HardParameterMap}
\end{figure}

Unlike alternative sets of the model coefficients from Table~\ref{CoefficientSet}, choosing one or another shape of injection function causes notable changes in the topology of the allowed  regions in the parameter space. Harder injection with $p=2.3$ results in  (somewhat counter-intuitively) 
a   best-fit solution with less pronounced KN effects and a smaller rate of secondary pair production {(smaller compactness)}. 
The net effect (see Fig.~\ref{HardParameterMap}) is a shift of the lower bound on the shock's Lorentz factor to larger values $\Gamma \gtrsim 180$, that is again set by the observed optical flux. The best-fit region shifts to even larger Lorentz factor $\Gamma \simeq 220$ and moves away from the point which matches the optical flux, although a global fit from optical band to TeV band with reasonable goodness is still possible.

A softer injection with $p=2.7$ leads, on the contrary, to a best-fit solution with more pronounced KN corrections and a larger rate of secondary pair production. 
As a result, it tends to produce SEDs with too soft TeV spectra and the region of fits with reasonable goodness shrinks to a small spot between $\Gamma \simeq 135$ and $\Gamma \simeq 175$, that again encompasses the solution which matches the observed optical flux. There is also a region of lower, but still satisfactory, goodness of fit, that extends towards much larger $\Gamma$ values and is a remnant of the valley-shaped region of allowed solutions found for $p=2.5$ and $p=2.3$.
{At $\Gamma \gtrsim 460$ there is another local minimum of $\chi^2$, which coincides with a good fit to the optical flux. However, this minimum is not as deep as the minimum at low $\Gamma$ and in addition it corresponds to unreasonably small external density.}

\begin{figure}
    \centering
 \includegraphics[width=1.0\columnwidth]{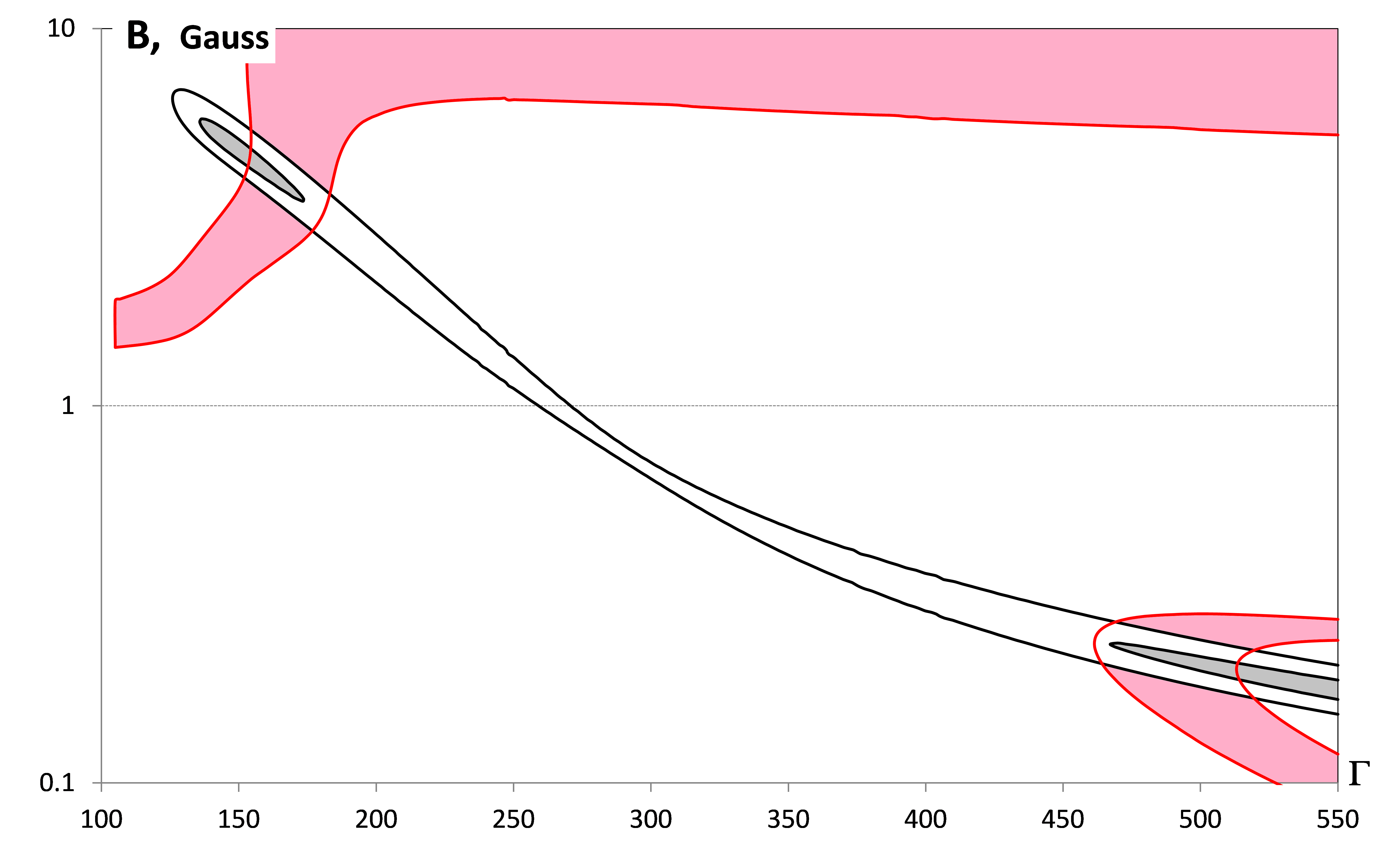}
    \caption{The regions of acceptable fits ($\chi^2 < 16$, light grey)  in $B$ -- $\Gamma$ parameter plane for soft injection with $p=2.7$. 
    An additional contour line indicates a region of mediocre fits ($\chi^2 < 25$, unshaded).
    The region where calculated optical flux is $0.7 \div 1.35$ of the observed value (covering both systematic and statistical uncertainties) is colored light red. Note that the overall goodness of  fit is lower here than in the other cases.  
    }
      \label{SoftParameterMap}
\end{figure}

\begin{figure}
    \centering
 \includegraphics[width=1.0\columnwidth]{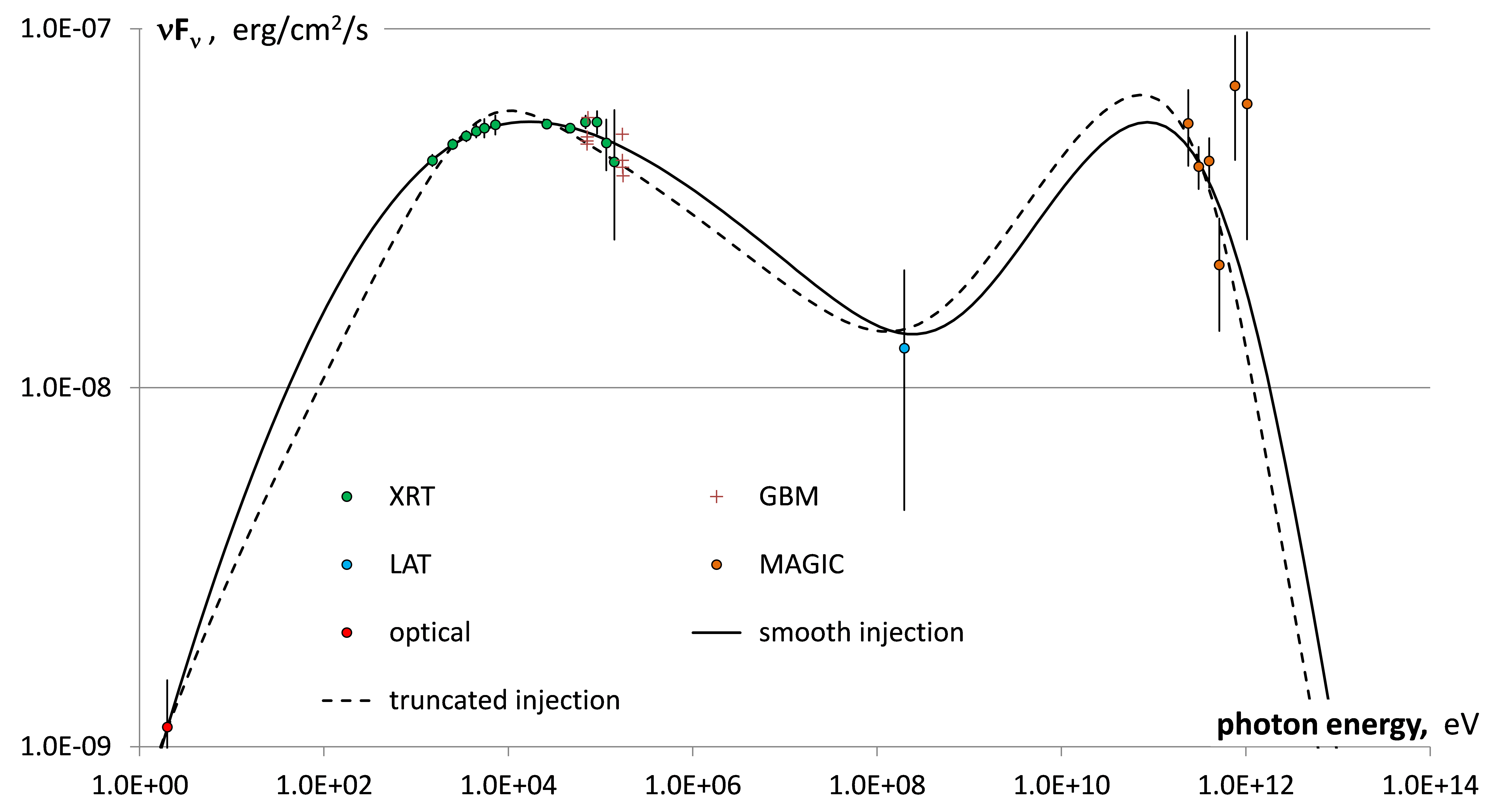}
    \caption{SED fits for smooth and truncated injection. Solid line --- SED fit for smooth injection with best $\chi^2$ in X-ray, GeV and TeV bands under additional requirement that it exactly fits the optical band flux ($\Gamma = 190$, $B = 5.21$~G {($\epsB \simeq 6.8 \times 10^{-3} / E_{53.5}$),} $\Ksc \simeq 13.4$, $\gamma_\mathrm{b} \simeq 5370$). Dashed line ---  SED fit for truncated injection  ($\Gamma = 158$, $B = 6.21$~G {($\epsB \simeq 3.2 \times 10^{-3} / E_{53.5}$),} $\Ksc \simeq 27.1$, $\gamma_\mathrm{m} \simeq 21000$), {which also matches the optical flux}. 
    Both fits use $p=2.5$ and \SPN{} that correspond to the ISM case.}
      \label{truncated_vs_smooth}
\end{figure}

Finally, we 
look for truncated-injection SED fit, commonly used in most afterglow studies.  
Here we also find a global, from optical to TeV, fit obtained with $p=2.5$.
The parameters of truncated-injection fit (see Fig.~\ref{truncated_vs_smooth}) are not much different from those of smooth-injection fit (note that for $p=2.5$ the average Lorentz factor of injected electrons is $\langle \gamma \rangle = 3 \, \gamma_\mathrm{m}$ and $\langle \gamma \rangle = 6 \, \gamma_\mathrm{b}$). 
 Both the smooth-injection and the truncated-injection SEDs
produce excellent fits to the observed optical, X-ray, GeV and TeV fluxes and look similar in their IC part (TeV data points). The most prominent difference is overly curved shape of the  synchrotron (X-ray) peak owing its origin to the sharp cut-off in truncated injection function.

\subsection{Late time SED and evolution of the parameters}
\label{sec:pivotal-late-time}
\begin{figure}
    \centering
 \includegraphics[width=1.0\columnwidth]{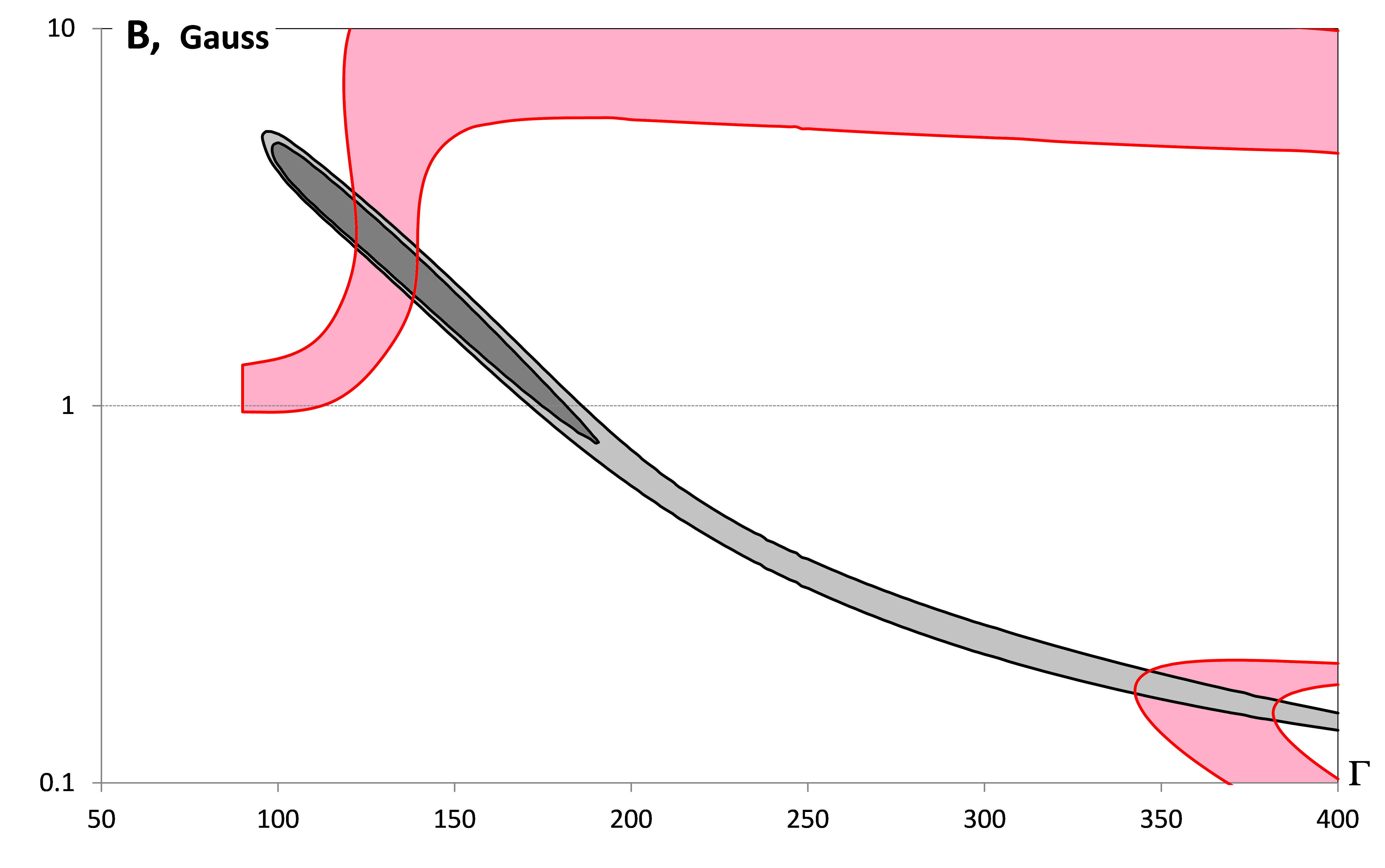}
 \includegraphics[width=1.0\columnwidth]{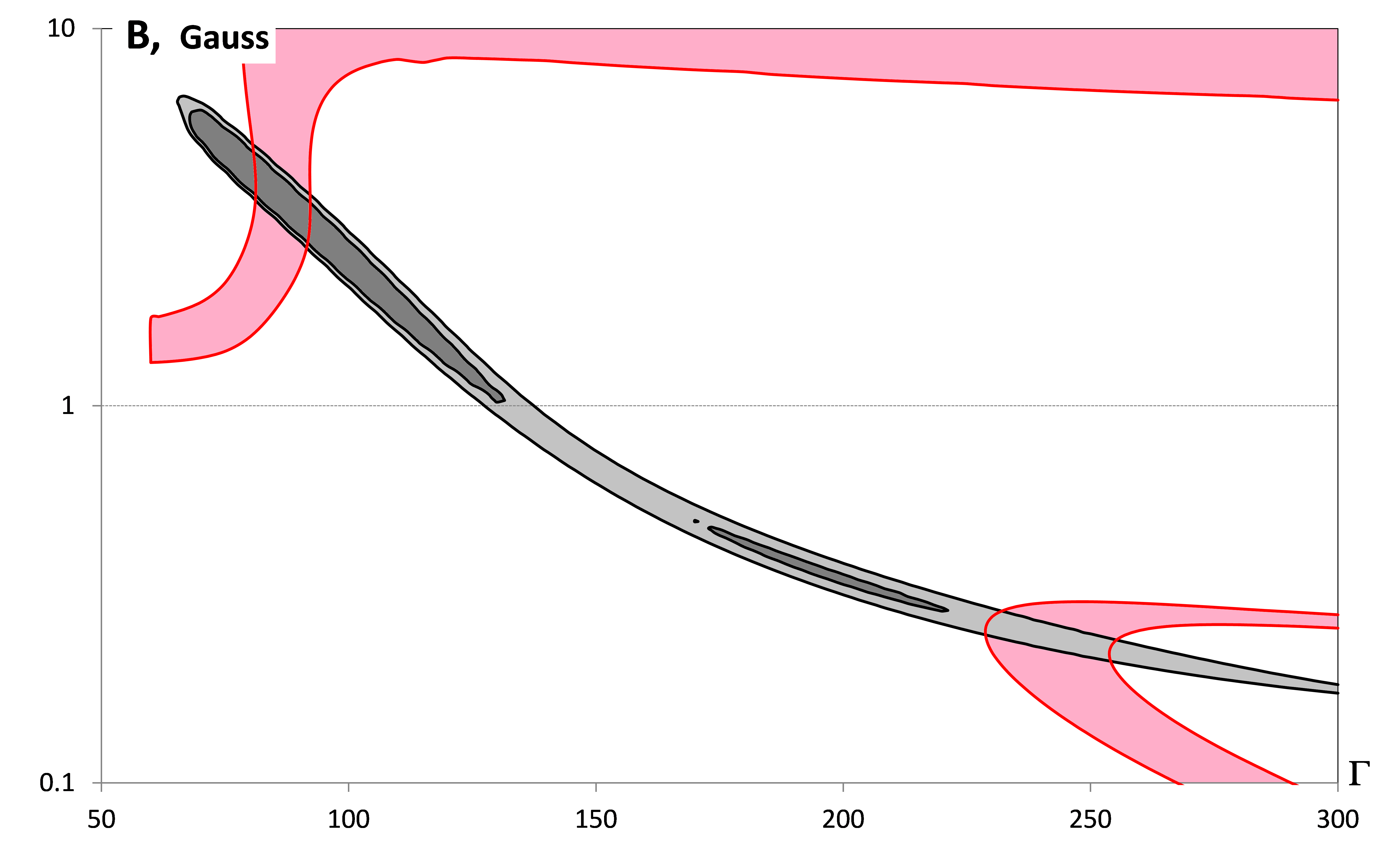}
    \caption{The SSC parameter plane at $t_\mathrm{obs} = 145$~s for wind density profile (top) and {ISM (constant  density)} profile (bottom). 
    The contour lines correspond to {$\chi^2 = 12$ (dark grey) and $\chi^2 = 16$ (light grey).}
    The region where calculated optical flux is $0.7 \div 1.35$ of the observed value (covering both systematic and statistical uncertainties) is colored light red. 
    }
      \label{LateParameterMap}
\end{figure}

\begin{figure}
    \centering
 \includegraphics[width=1.0\columnwidth]{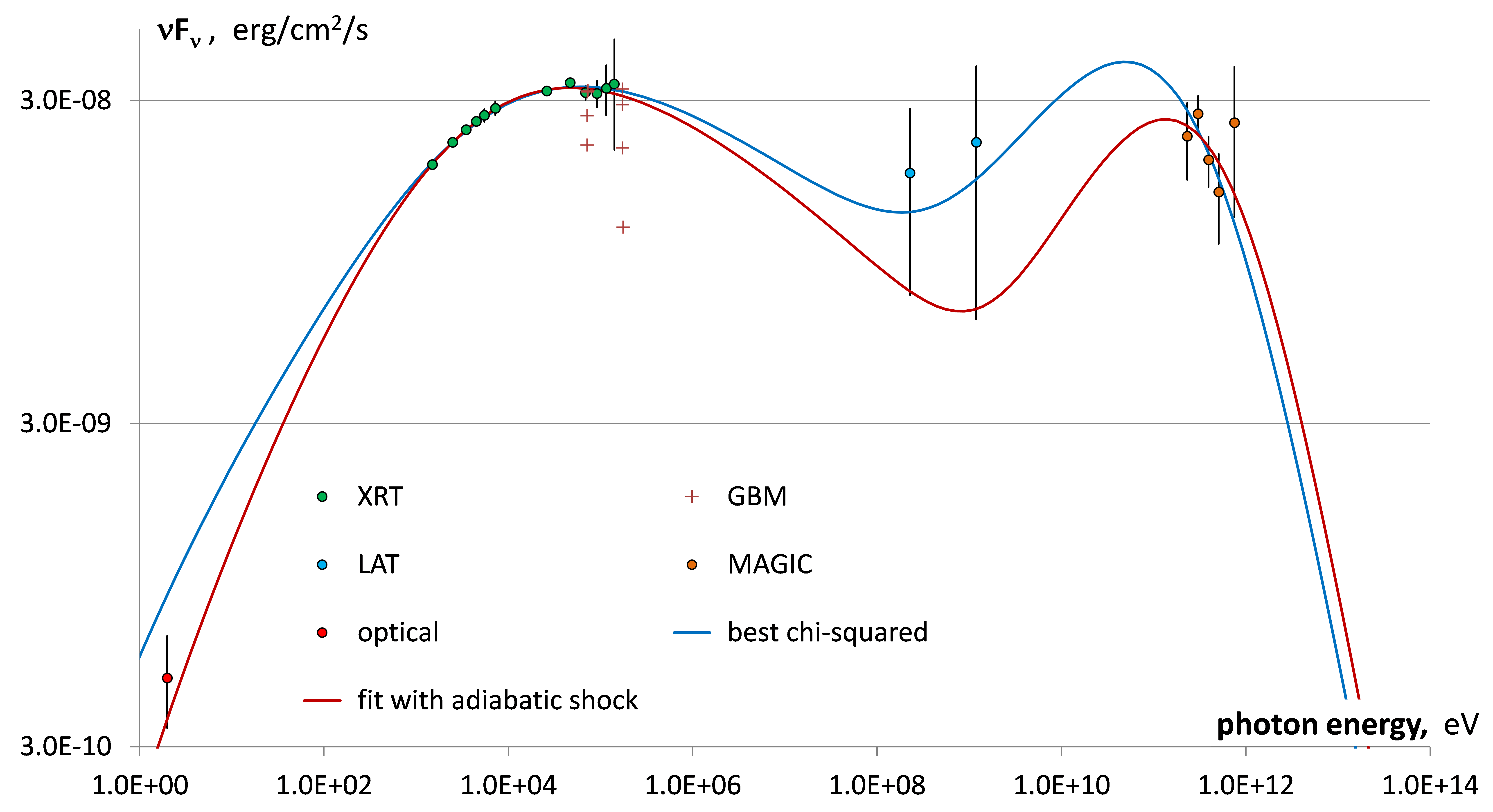}
 \includegraphics[width=1.0\columnwidth]{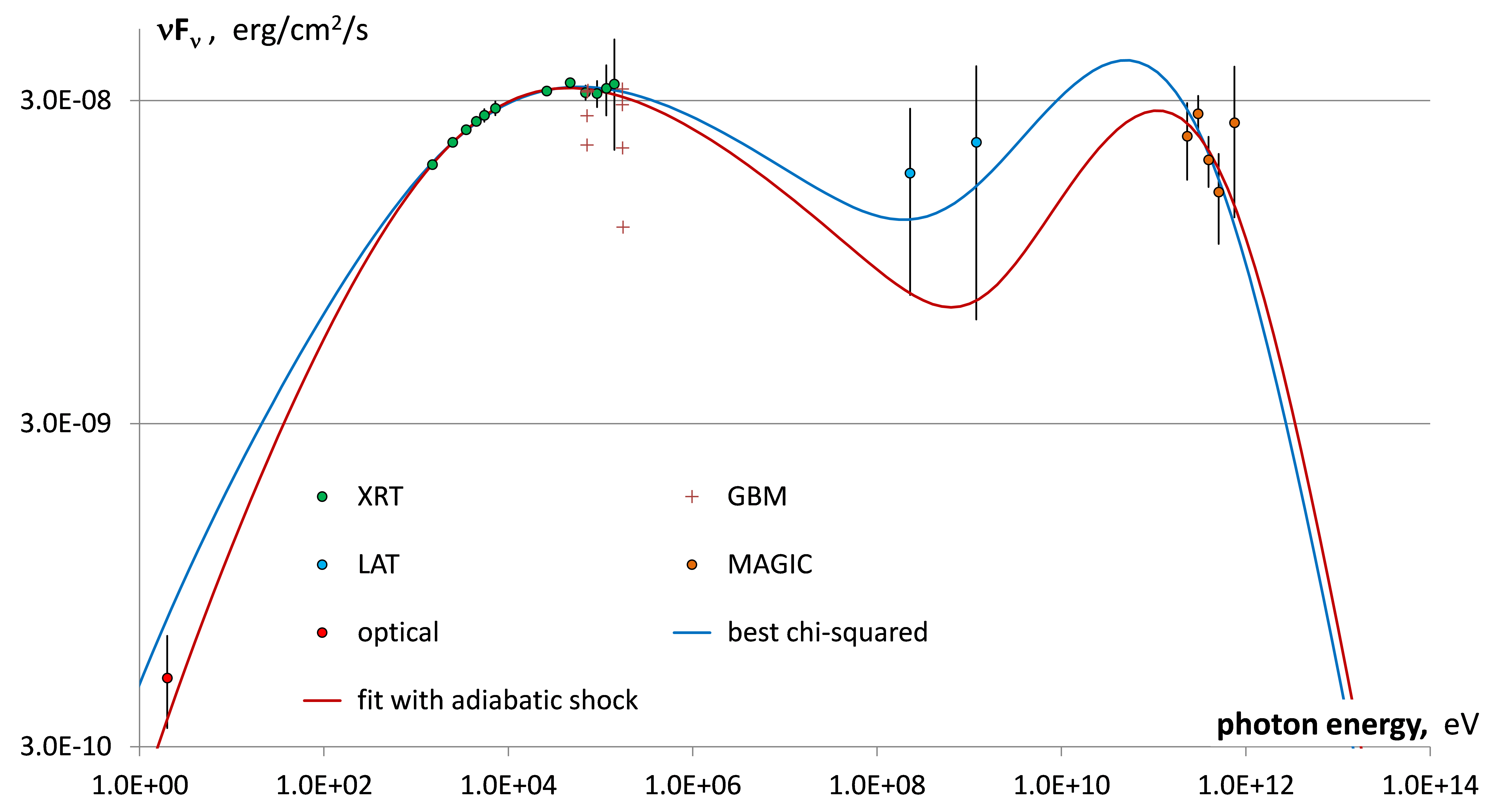}
    \caption{SED fits at $t_\mathrm{obs} = 145$~s. Blue lines --- best $\chi^2$ SED fits to X-ray, GeV and TeV data points. 
    {
    Red lines --- SED fits that correspond to the same circumburst density parameters for both the early and the late observations: progenitor's mass loss rate ($1.4 \times 10^{-6} V_\mathrm{w,3000} E_\mathrm{53.5}\  M_{\sun}$/yr) or fixed ISM density ($2\, E_\mathrm{53.5}\ m_p$/cm$^3$).
    {\bf Top}:  wind density profile. 
    Best fit parameters are $\Gamma = 115$, $B = 3.46$~G {($\epsB \simeq 2.13 \times 10^{-3} / E_\mathrm{53.5}$),} 
    $\Ksc \simeq 55.9$, $\gamma_\mathrm{b} \simeq 15000$, resulting in $\chi^2 \simeq 5.65$ for 19 {data points} (not counting the optical point) with four parameters.
    Fixed progenitor's mass-loss rate fit parameters are $\Gamma = 143$, $B = 2.02$~G {($\epsB \simeq 2.70 \times 10^{-3} / E_\mathrm{53.5}$),} 
    $\Ksc \simeq 35.7$, $\gamma_\mathrm{b} \simeq 16700$, resulting in $\chi^2 \simeq 7.57$ for 19 {data points} (not counting the optical point) with four parameters.
    {\bf Bottom}:   ISM density profile. 
    Best fit parameters are $\Gamma = 79.1$, $B = 4.28$~G {($\epsB \simeq 2.14 \times 10^{-3} / E_\mathrm{53.5}$),}
    $\Ksc \simeq 59.0$, $\gamma_\mathrm{b} \simeq 13300$, resulting in $\chi^2 \simeq 5.64$ for 19 {data points} (not counting the optical point) with four parameters.
    Fixed ISM density fit parameters are $\Gamma = 91$, $B = 3.11$~G {($\epsB \simeq 2.62 \times 10^{-3} / E_\mathrm{53.5}$),}
    $\Ksc \simeq 40.9$, $\gamma_\mathrm{b} \simeq 14400$, resulting in $\chi^2 \simeq 6.74$ for 19 {data points} (not counting the optical point) with four parameters.
    }
    }
      \label{LateTimeSEDs}
\end{figure}

We apply the same procedure for the later-time data set, that covers interval from 110 to 180~s after the trigger, attempting to find the best fit with instantaneous SED calculated for $t_\mathrm{obs} = 145$~s. The resulting best (minimal) SED $\chi^2$ values achieved in our parameter space scans are shown in Fig.~\ref{LateParameterMap} as function of $\Gamma$ and $B$.
The topology of allowed regions in the parameter space is similar to those obtained with earlier-time data set (Figs.~\ref{Fig.early_parameters_space} and \ref{ISM_parameter_map}) and again, the best $\chi^2$ solutions based on X-ray, GeV and TeV data points are close to the solutions that in addition reproduce the observed optical flux. The corresponding SEDs are shown in Fig.~\ref{LateTimeSEDs}. Thus, a later-time global SED fit from optical to TeV is possible. Like in the earlier data set, a second (slow cooling) solution exists that provides optically consistent fit. Again, it has worse goodness of fit, 
it corresponds to larger $\epsB$ {and $\epse$} values, and it implies unreasonably small density of the circumburst medium.

There are several points worth mentioning when we compare earlier-time fits to later-time ones. 
First,  the later-time fits clearly {favor} smaller shock's Lorentz {factor ---} as expected, the shock slows down. 
Comparing the best $\chi^2$ solutions for early and late time we find that they imply fast temporal evolution of $\Gamma$, faster than both wind and ISM cases for a constant-energy blast wave. This may be considered as an evidence of significant radiative energy loss, but statistical support for this conclusion is weak --- it is possible to choose different  pairs of solutions that obey  the deceleration law of an adiabatic shock that have good $\chi^2$ values. 
The wind density profile solutions that we find require reasonable mass loss rate and we choose them as reference solutions.
Their parameters are summarized in Table~\ref{ParametersSummary}
and the corresponding SEDs are shown in Figs.~(\ref{fits_at_early_time}) and (\ref{LateTimeSEDs}).

\begin{table*} 
\caption{Summary of afterglow parameters found for a combined best fit solutions at two moments of {time assuming} adiabatic evolution of the shock wave. 
In the table, parameters obtained with a wind density profile are followed in parentheses by parameters obtained with an 
ISM density profile.
Note that the parameters below the double line depend on specific choice of the shock's kinetic energy.
}
\label{ParametersSummary}
\begin{tabular}{|l|c|c|}
\hline
parameter  & $t_\mathrm{obs} = 90$~s  & $t_\mathrm{obs} = 145$~s \\
\hline
$\Gamma$ & $161$ ($109$) & $143$ ($91$) \\
\hline
$B$ & $4.4$~G ($5.7$~G) & $2.0$~G ($3.1$~G) \\
\hline
$\Ksc$ & $20$ ($21$) & $36$ ($41$) \\
\hline
$\gamma_\mathrm{b}$ & $6500$ ($5700$) & $16700$ ($14400$) \\
\hline
$p$ & $2.5$ & $2.5$ \\
\hline
\hline
$E_\mathrm{kin}$ & $3 \times 10^{53}$~erg & $3 \times 10^{53}$~erg \\
\hline
$\epsB$ & $0.0061$ ($0.0062$) & $0.0027$ ($0.0026$) \\
\hline
$\epse$ & $0.12$ ($0.13$) & $0.096$ ($0.107$) \\
\hline
$\dot{M}$ (wind) & $1.4 \times 10^{-6}  \frac{V_w}{3000km/s}  M_{\sun} /yr$ & $1.4 \times 10^{-6}  \frac{V_w}{3000km/s}  M_{\sun} /yr$ \\
\hline
$n$ (ISM) & $2$~cm$^{-3}$ & $2$~cm$^{-3}$ \\
\hline
\end{tabular}
\\
\end{table*}

We cannot choose two SED fits which are in good agreement with the optical observations while also obeying the adiabatic shock deceleration law, either with wind or ISM density profile.  
A decrease of the shock energy with time due to radiative losses relaxes somewhat  this discrepancy so that the remaining difference in the optical-band fluxes is $\sim 20\%-30\%$.  

Based on our model SEDs we find that from $t_\mathrm{obs}=90$~s and to infinity  $\sim 1.3\times 10^{53}$~erg was radiated, setting a lower limit for the shock energy.
The best-fit parameters
hint at a significant energy decrease between 90 and 145 seconds, so we
are not far from this lower limit.
Therefore, we will use $E_\mathrm{kin}=3\times 10^{53}$~erg as a reference value {denoting it as $E_{53.5}$}.
The best fit solutions that are consistent with adiabatic evolution from the early to the late epoch correspond to 
mass-loss rate 
$1.4\times 10^{-6} V_\mathrm{w,3000} E_{53.5} \ M_{\sun}$/yr (here $V_\mathrm{w,3000}$ is the wind velocity in units 3000~km/s)
in the wind case and to density $n \simeq 2 \, E_{53.5} \ $/cm$^3$  for ISM.
Neither wind nor ISM is clearly preferred over  the other  and we need additional arguments to tell the two cases apart. 

\begin{figure}
    \centering
 \includegraphics[width=1.0\columnwidth]{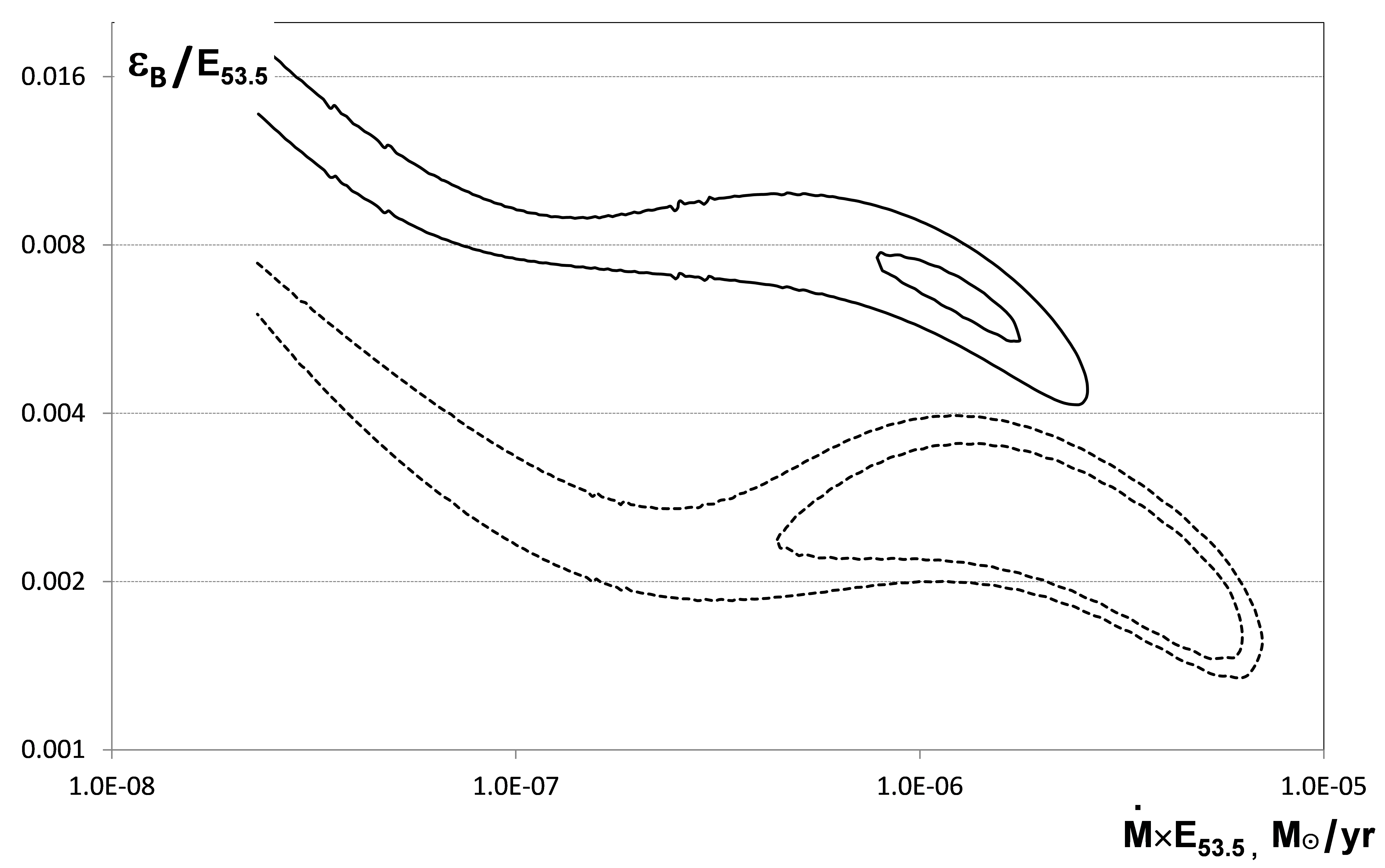}
 \includegraphics[width=1.0\columnwidth]{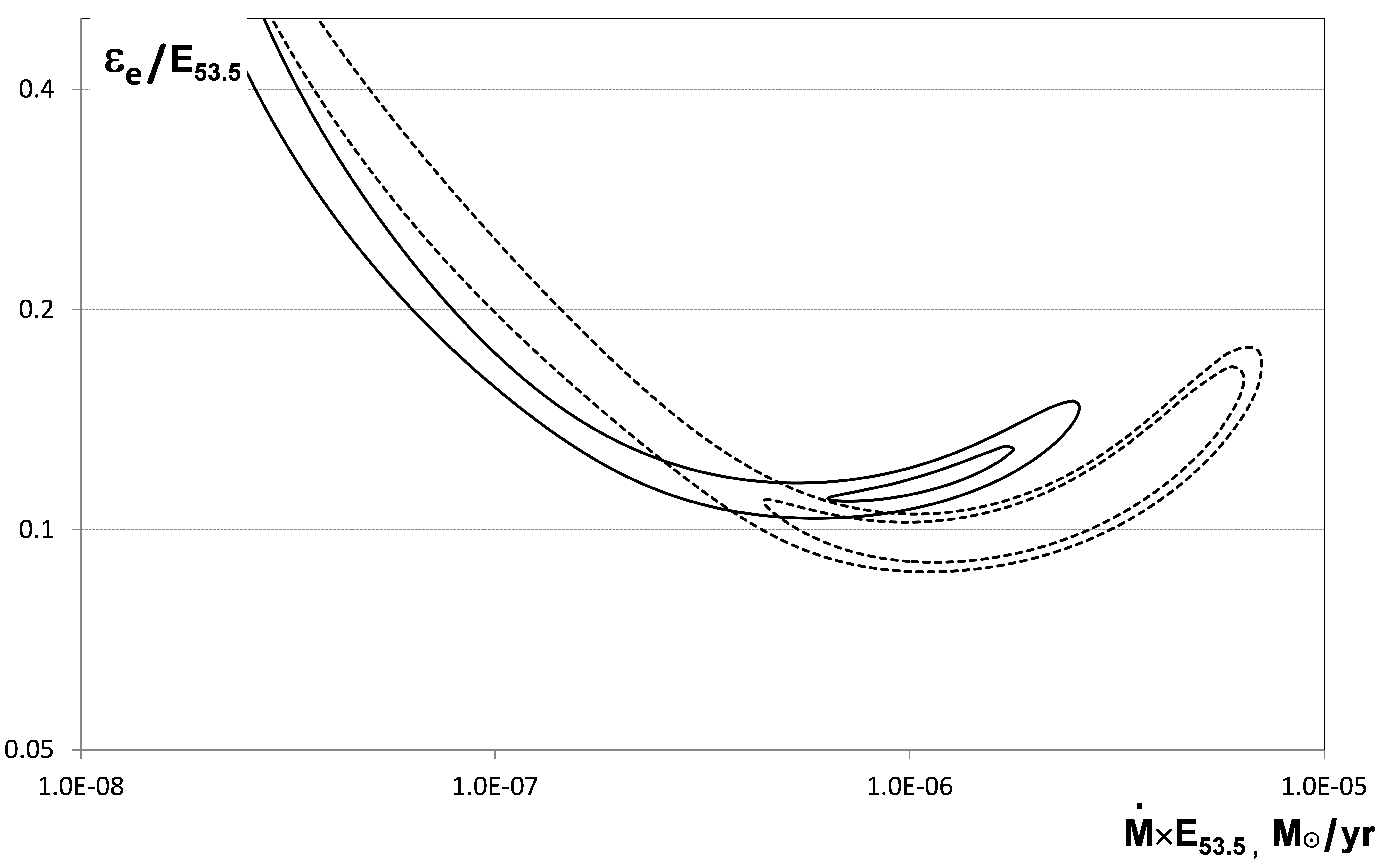}
    \caption{
   {
   Regions of acceptable parameters (based on X-ray, GeV and TeV data points) shown in $\epsB$ -- $\dot{M}$ (top) and  $\epse$ -- $\dot{M}$ (bottom) coordinates for $t_\mathrm{obs} = 90$~s (solid lines) and $t_\mathrm{obs} = 145$~s (dashed lines). 
    The mass loss rate is calculated assuming wind velocity $V_w = 3000$~km/s.
    The contour lines mark regions of good fits ($\chi^2 < 12$) and regions of acceptable fits ($\chi^2 < 16$).
    Note that while the allowed regions for $\epse$ overlap, those for $\epsB$ don't. 
    Radiative losses decrease later-time shock's energy and hence the regions of acceptable parameters calculated for later time shift slightly upwards and to the left.}
    }
      \label{ParametersEvolution}
\end{figure}

Second, the fit parameters clearly show that the average Lorentz factor of the injected electrons increases with time, in contradiction with the usual assumption  $\gamma_\mathrm{b} \propto \Gamma$. 
As the earlier- and later-time fits are consistent with $\epse \approx const$ {(see lower panel in Fig.~\ref{ParametersEvolution})},  this implies that the fraction of accelerated electrons decreases while the shock expands.
Such a strong evolution of the fraction of accelerated electrons is  a natural feature of the pair-balance model \citep{PairBalance}.
Interestingly the fraction of self-absorbed high-energy photons is the same ($\simeq 10\%$) 
in both the late and early solutions,  in line with expectation of the pair balance model. 

Third, the earlier- and later-time fits exclude constant $\epsB$ for an adiabatic blast wave {(see upper panel in Fig.~\ref{ParametersEvolution})}.
However the condition $\gamma_\mathrm{b} \simeq \left( B_\mathrm{cr}/B \right)^{1/3}$ holds again in agreement with the pair balance model.  

Forth, 
both early and late time  solutions are very fast cooling (only $\simeq 10\%$ of the total injected power  remain in the electrons). The system is relatively strongly absorbed ($\simeq 10$\% 
of the total power of injected electrons is eventually absorbed and comes out as secondary pairs at lower energies).  A significant fraction of the radiation power is in the IC component (decreasing from $\simeq 40\%$ at early time to $\simeq 30\%$ at late time).

\section{Comparison with other works}
\label{sec:comparison}

Soon after the discovery of GRB~190114C was announced, several papers analyzed the afterglow parameters yielding significantly different results. We compare in this section some  of these results with the results obtained with the code used  here.

The first study of GRB~190114C afterglow parameters \citep{DerishevPiran2019} was based on MAGIC detection announcement and data from the corresponding GCN messages. Since the data was preliminary and incomplete, the study did not go beyond analytic order-of-magnitude estimates, yielding however some non-trivial results. Those are: the shock's Lorentz factor $\Gamma \simeq 100$ at $t_\mathrm{obs} = 70$~s, the 
effective Lorentz factor of emitting electrons $\gamma_\mathrm{eff} \simeq 10^4$, 
radiation in the fast cooling regime with Comptonization on the border of  KN  regime, unexpectedly large magnetic energy fraction $\epsB \simeq 0.1$ likely reduced by a factor of several provided the IC peak is strongly suppressed by internal two-photon absorption. 
Qualitatively, these results are supported by much more elaborate analysis in the present paper, although the numbers are somewhat revised --- partly because the initially reported lower limit on TeV flux appeared to be more than two times smaller than the actual value and partly because of combined action of KN corrections  and internal two-photon absorption, whose influence is possible to evaluate only numerically.

\begin{figure}
    \centering
 \includegraphics[width=1.0\columnwidth]{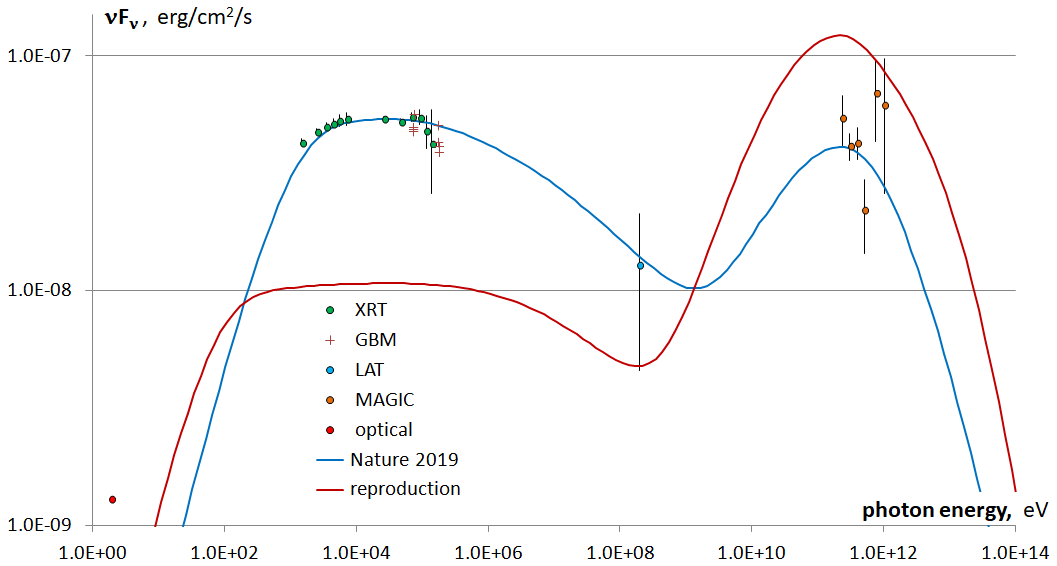}
 \includegraphics[width=1.0\columnwidth]{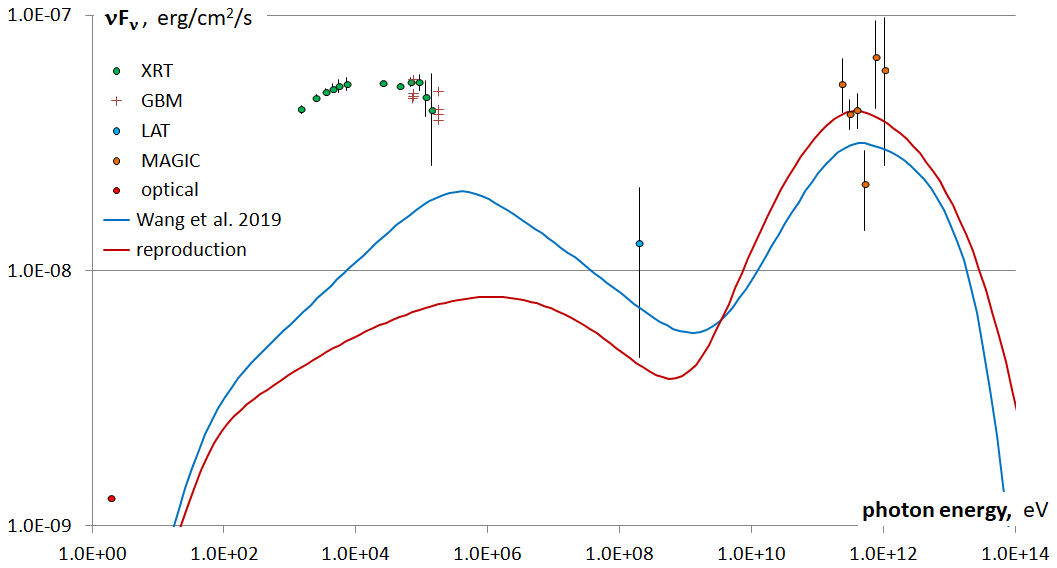}
 \includegraphics[width=1.0\columnwidth]{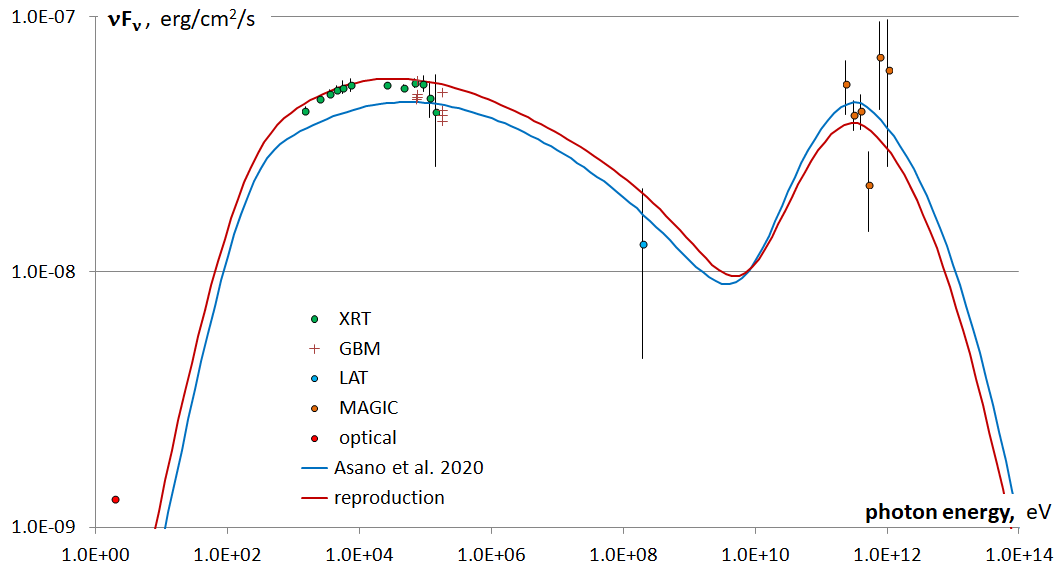}
    \caption{{SEDs from \protect\cite{MAGIC_Nature_fit} (top panel, blue curve), \protect\cite{Wang_etal_SED} (middle panel, blue curve), and \protect\cite{Asano_etal_SED} (bottom panel, blue curve) and their numerical reproduction (red curves). The parameters of the emitting zone for each of the fits are given in the text.}
    }
      \label{SEDcomparison}
\end{figure}

We compare the {SEDs calculated by} \cite{MAGIC_Nature_fit,Wang_etal_SED,Asano_etal_SED} to {SEDs} calculated using our code with the same parameters. All three use ISM density profile with the \SPN{} and truncated injection. When trying to reproduce their results we use the parameters presented in the papers and adopt the same assumptions about the density profile, the coefficients, and the shape of the injection function.

The model  parameters given in the discovery paper \cite{MAGIC_Nature_fit} are:  external density $n = 0.5$~cm$^{-3}$, 
shock's kinetic energy $E_\mathrm{kin} = 8 \times 10^{53}$~erg,
observation time $t_\mathrm{obs} = 90$~s (in the middle of the quoted time interval 68 -- 110~s),
$\epse = 0.07$,
$\epsB = 8 \times 10^{-5}$,
and truncated injection with index $p=2.6$.
From this set of parameters we calculated
$\Gamma = 245$ (to avoid confusion let us remind that in our notations  $\Gamma$ is the shock's front Lorentz factor, that is $\sqrt{2}$ times larger than the bulk material Lorentz factor immediately behind the shock),
$B=0.43$~G,
and $\gamma_\mathrm{m} = 8350$ (following {Eq.~(\ref{avg_gamma}) with $\xi_e=1$}). 
With the above parameters our numerical results (see top panel in Fig.~\ref{SEDcomparison}) strongly deviate from the SED presented in \cite{MAGIC_Nature_fit} and we could not come even to a qualitative agreement by varying parameters around these values. The main reason for disagreement is very large value of $\Ksc$ derived in that work. This causes the IC peak flux to be about an order of magnitude higher than the synchrotron peak flux  even though we take into account both KN corrections  and internal two-photon absorption that  reduce  the {high-energy}  emission.

{One of the first attempts to explain the observations was by \cite{Wang_etal_SED} who tried to model only the TeV emission of this source. }
The model parameters  are:
external density $n = 0.3$~cm$^{-3}$, 
shock's kinetic energy $E_\mathrm{kin} = 6 \times 10^{53}$~erg,
observation time $t_\mathrm{obs} = 90$~s (in the middle of the quoted time interval 50 -- 150~s),
$\epse = 0.07$,
$\epsB = 4 \times 10^{-5}$,
and truncated injection with index $p=2.5$.
From this set of parameters we calculated
$\Gamma = 252$, 
$B=0.24$~G,
and $\gamma_\mathrm{m} = 7630$. 
With the above parameters the agreement between our numerical results (see middle panel in Fig.~\ref{SEDcomparison}) and the SED presented in \cite{Wang_etal_SED} is only qualitative, with the largest deviation around the synchrotron peak. 
The model proposed by  \cite{Wang_etal_SED}    attributes the X-ray peak to a different origin. Consequently, both the original SED and -- especially -- its numerical reproduction fall short of the observed X-ray flux with this set of parameters. 

The  model parameters given in \cite{Asano_etal_SED} are: 
external density $n = 0.3$~cm$^{-3}$, 
shock's kinetic energy $E_\mathrm{kin} = 4 \times 10^{53}$~erg,
observation time $t_\mathrm{obs} = 80$~s,
$\epse = 0.1$,
$\epsB = 1 \times 10^{-3}$,
and truncated injection with index $p=2.3$.
From this set of parameters we calculated
{$\Gamma = 251$}, 
$B=1.2$~G,
and $\gamma_\mathrm{m} = 7520$. 
Additionally, \cite{Asano_etal_SED} assume escape time for photons 
that is 6 times smaller than  the  lifetime 
of the  emitting electrons. This corresponds, in our notations, to a geometrical factor $\Lambda = 1/6$ (see Eq.~\ref{photon_equation}) {that} is hardly possible for a slab geometry\footnote{\cite{Asano_etal_SED} argue that  the
electrons move across the emitting zone with the {hydrodynamical}  speed $c/3$, while
photons escape from the center of the emitting zone with the speed $c$ along the shock's normal,
i.e., their escape time is 6 times smaller. However, first, the emitting zone's boundary is not stationary, but it also moves away with the speed $c/3$. Second, photons that move at an angle to the shock's normal escape slower and in a slab geometry the effective photon lifetime averaged over different propagation angles  diverges logarithmically unless one takes into account the slab's expansion.}. 
The  SED calculated  with the above parameters  (using $\Lambda = 1/6$) is in  a reasonable agreement with the results of \cite{Asano_etal_SED} (see bottom panel in Fig.~\ref{SEDcomparison}). The difference between the original SED and its reproduction exceeds the anticipated precision of the code we used, and it cannot be attributed to inclusion of internal two-photon absorption into our code. However, the results can be brought to an agreement with a minor change in the parameters. Extrapolating to the usual $\Lambda = 1$ assumption, one needs about 6 times larger $\epsB$ (about 2.5 times stronger magnetic field) to obtain a {similar SED}. With this correction the corresponding  solution 
is not far from the valley of good fits that we find, although it is far from a considerably better fit found in the low-$\Gamma$ end of this valley.

\begin{figure}
    \centering
 \includegraphics[width=1.0\columnwidth]{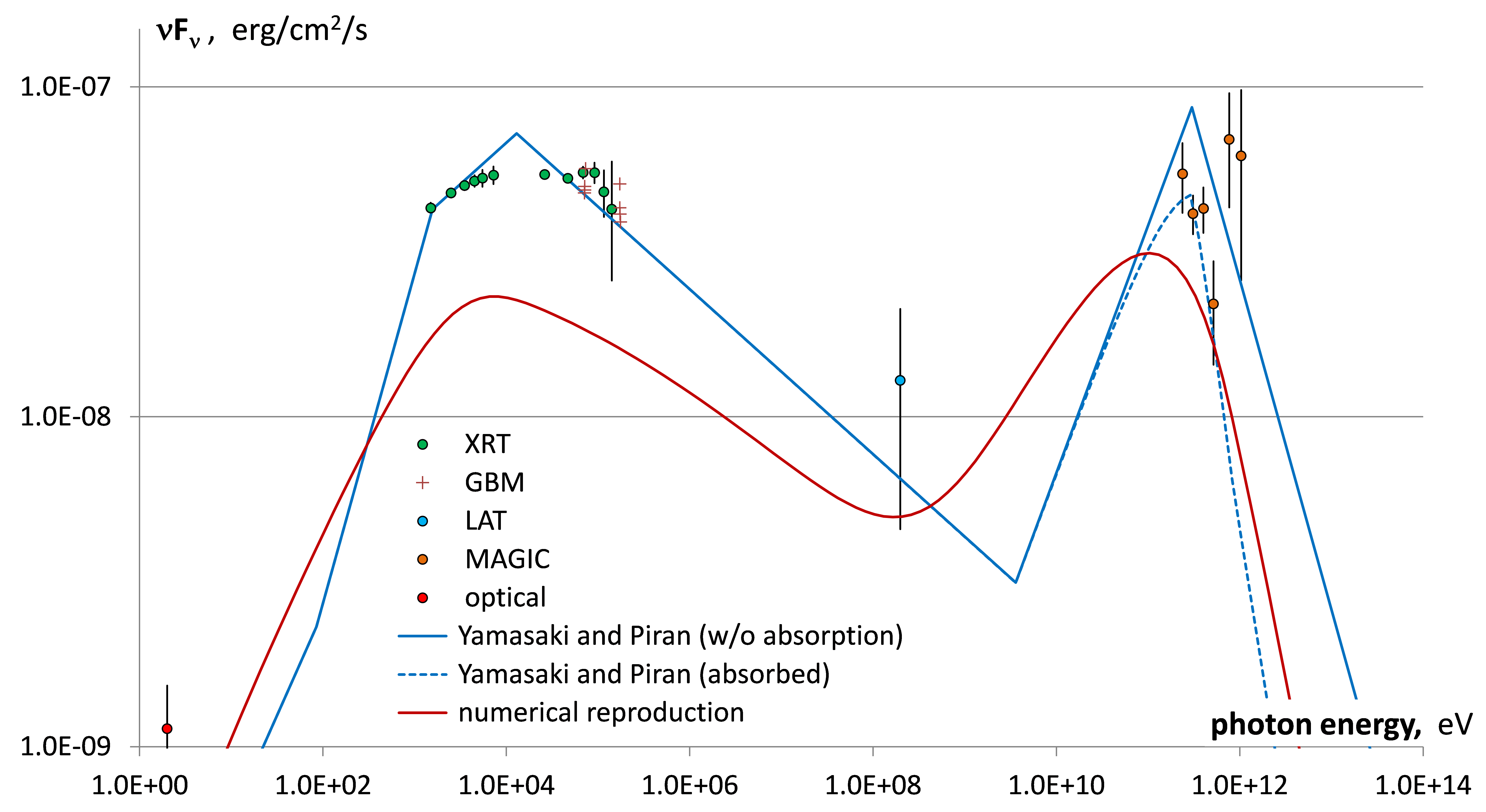}
    \caption{
    A comparison of  the observations  with numerical {(red curve)}   
    and  analytic {(blue curve)} {SEDs calculated for $t_\mathrm{obs}=90$~s}. The analytic model \citep{Yamasaki+21}
    follows a  modified \protect\cite{NakarAndoSari2009} formulation and uses the same parameters as the numerical {one.} Fit parameters are: 
shock's kinetic energy $E_\mathrm{kin} = 3 \times 10^{53}$~erg, 
constant-density external medium with $n = 12$~cm$^{-3}$, 
shock's front Lorentz factor $\Gamma = 146$, 
$B = 4.6$~G  ($\epsB = 1.1 \times 10^{-3}$),
$\epse = 0.038$  ($\Ksc = 35.6$), $\gamma_\mathrm{m} = 20300$,
truncated injection with index $p=2.5$ and \SPN{} that correspond to an ISM.}
      \label{SEDcomparison_analytic}
\end{figure}

Figure~\ref{SEDcomparison_analytic}  depicts a comparison of the numerical SED with 
\SPN{} and a truncated injection   and the
analytic SED based on \cite{NakarAndoSari2009} using the same parameters. 
It is important to note here that in obtaining this SED \cite{Yamasaki+21} assumed that the KN decline in the scattering cross-section is abrupt and occurs at photon energy $0.2 m_e c^2$ (in the electron's frame) rather than at energy $m_e c^2$ as in the original \cite{NakarAndoSari2009} formulation.
This modification takes into account the fact that KN effects 
begin to take place for electron energies lower than $m_e c^2$.
Self absorption  of {high-energy}  photons due to pair production with {low-energy} photons is not taken into account in this analytic calculation. This leads to some excess of {high-energy} photons {as well as} lack of {low-energy} photons that in the numerical calculation are produced mostly by these secondary pairs.  

\section{Comparison with the Pair Balance Model}
\label{sec:Pair.balance}

\begin{figure}
    \centering
\includegraphics[width=1.0\columnwidth]{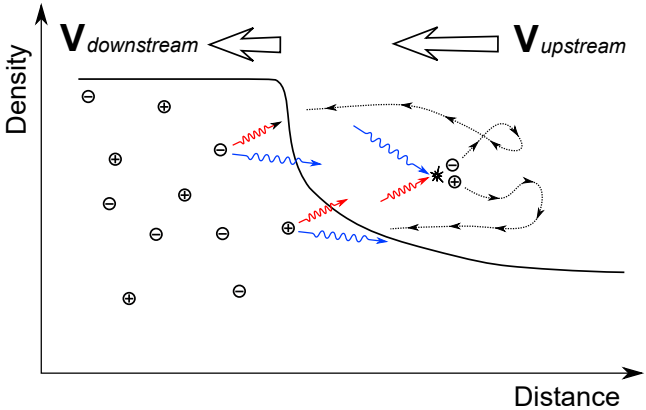}
    \caption{
    A sketch of the modified relativistic shock in the pair-balance model. High-energy IC photons interact with low-energy synchrotron photons ahead of the shock to produce electron-positron pairs, which are intercepted by the upstream magnetic field. The pairs transfer momentum to the upstream matter accelerating and  compressing it, so that the velocity jump at the shock front is greatly reduced. In the downstream, the pairs produce synchrotron and IC radiation. In the upstream they build up the magnetic field.}
      \label{fig:pairbalance}
\end{figure}

Parameters of the best fit solution, that we find for GRB~190114C, are such that a significant fraction of radiated power is absorbed within the emitting zone by annihilation of  high-energy (IC) with low-energy (synchrotron) photons producing secondary electron-positron pairs. The very same process inevitably takes place in the upstream region of the shock. A consistent description of GRB afterglows must, therefore,  take this into account and suggests using the pair balance model \citep{PairBalance}. In this model the pairs produced in the upstream deposit energy and momentum and modify both the hydrodynamic flow and the magnetic fields in this region (see Fig.~\ref{fig:pairbalance}). Thus, the shock propagates into an already perturbed upstream that contains a significant fraction of high energy pairs as well as seed magnetic field. 

This leads to several important  differences from the regular diffuse shock acceleration. First 
the high energy photons serve as the neutral agents that transfer energy across the shock from the downstream to the upstream region, within the  converter {acceleration} mechanism \citep{converter}. 
The newly created pairs gain energy by a factor up to  $\sim \Gamma^2$
once they cross the shock. In the downstream these accelerated leptons emit both low energy (synchrotron) and high energy (inverse Compton) photons. Some of the  fresh higher energy  photons  create comparable higher energy pairs once they annihilate in the upstream. These will gain another factor  up to $\sim \Gamma^2$ upon crossing the shock. 
In a converter acceleration the energy of the accelerated particles would diverge unless something turns down the energy transfer across the shock.  Here, 
as the  high energy photons are produced in the downstream by inverse Compton, 
the process  becomes inefficient once the KN  regime is reached. As the low energy photons are produced via synchrotron {emission,} this sets a natural limit on the typical electrons' Lorentz factor, $\gamma_\mathrm{b}$, and the magnetic field $B$ within the downstream region, such that $\gamma_\mathrm{b} \simeq \left( B_\mathrm{cr}/B \right)^{1/3}$.

A second  important and essential  aspect of this model, 
that  doesn't have a direct effect on the spectrum discussed here,  is the decay of the magnetic field in the downstream, that takes place over a large scale.    As the seeds of the magnetic field were generated in the upstream over a scale of order of the optical depth for pair creation,  its decay won't be over a skin depth,  but rather on a scale comparable to this  macroscopic length.

Third, 
operation of the pair balance model {implies} that a fixed and non-negligible fraction of the high energy (IC) photons  annihilate in the upstream producing pairs by interaction with lower energy (synchrotron) photons. Since both the IC and the synchrotron photon field densities are similar in the upstream and downstream regions we expect that a comparable, constant and non-negligible  fraction of the IC photons will be self-absorbed within the downstream region as well. 

Forth,  a distinct component of high-energy leptons is injected into the upstream. This breaks the one-to-one relation  between the number of emitting leptons and the number of external electrons that the shock front has crossed (e.g. $4 \pi n  R_\mathrm{sh}^3/3$ for ISM). 
Put differently, this introduces $\xi_e \ne 1$ and relaxes  the common link Eq.~(\ref{avg_gamma}). 

As we have seen earlier, features
such as varying in time $\xi_e$, {increase of $\gamma_{\rm b}$ while the shock decelerates,} approximate equality of $\gamma_{\rm b}$ and  $\left( B_\mathrm{cr}/B \right)^{1/3}$, large (and constant) fraction of self-absorbed IC photons,
that emerged from the best fit to the observed data are {in full agreement with}  predictions of the pair balance model.  
On the other hand, some of these features (e.g. significant variation of $\xi_e$ and increasing  $\gamma_{\rm b}$ while $\Gamma$ decreases) are incompatible with common assumptions concerning {particle}  acceleration in shocks.  Those  that are compatible, like $\gamma_{\rm b} \simeq  \left( B_\mathrm{cr}/B \right)^{1/3}$, are not predicted.

\section{Implications}
\label{sec:summary}

Observations-wise, the prompt $\gamma$-rays and the low energy afterglow of   GRB~190114C  are similar to those of an ordinary burst with larger than average luminosity.  But this was a unique occasion when a GRB was detected in the sub-TeV range early in the afterglow and its simultaneous SED was measured from $\sim 1$~eV to  $\sim 1$~TeV. 
We used this opportunity to analyze in unprecedented detail the best-fit SED at two moments of time.
While it was generally believed that detection of both a low energy (synchroton) and a high energy (IC) component would {straightforwardly} reveal the conditions within the emitting region, it turns out that this wasn't the case and this task was quite challenging. 

Beyond obtaining the parameters of the emitting region in this burst our {goal was} to present a new systematic method of fitting SSC spectrum in the regime of   KN  correction {accompanied by self-absorption of IC photons,}  and to compare the conditions determined for this burst with afterglow modeling.  We find an excellent fit to the data all the way from optical to TeV. However,
our 
results suggest that afterglow parameters and their evolution in time deviate from common expectations.

To make sure that we are not missing any possible solution consistent with observations, we performed a systematic scan over 4-dimensional parameter space ($\Gamma$, $B$, $\Ksc$, and $\gamma_\mathrm{b}$) {that was possible thanks to good speed of a new one-zone SSC code \citep{newcode}, which we used}. This scan reveals a  valley of good fits --- an allowed region in the parameter space, which stretches along $\Gamma$ axis being relatively narrow in 3 perpendicular directions. For a fixed $\Gamma$, the valley's extent in $B$, $\Ksc$, and $\gamma_\mathrm{b}$ is dictated by observations of the IC (sub-TeV) component --- therefore, they are crucial in determining the emitting zone's parameters. Surprisingly, the allowed range of $\Gamma$ is limited by optical observations both from above and from below. The SEDs corresponding to the low-$\Gamma$ and the high-$\Gamma$ ends of this {range} 
match the observed optical flux in addition to fitting the X-ray, GeV and sub-TeV data points. Therefore, the entire spectrum from $~1$~eV to  $~1$~TeV is well reproduced by a single-zone model without invoking additional sources of low-energy photons. 

{The quality of the fit within the allowed parameter range is not uniform.}
The region at the low-$\Gamma$ end  has a  significantly better goodness of fit than the one at {high} $\Gamma$. Moreover, the 
high-$\Gamma$ end of the allowed parameters  corresponds to unreasonably low external density. Hence we consider the low-$\Gamma$ end as the  best fit solution.
Solutions at intermediate $\Gamma$ are allowed by the X-ray to TeV data but they  would require an additional source of optical emission to match the observations. Solutions with  higher or lower $\Gamma$ values {overproduce} the optical signal.  

With different sets of {model} coefficients (e.g. corresponding to either wind or ISM external density profile), the allowed region in the parameter space preserves its topology --- a relatively narrow valley that stretches along the $\Gamma$ axis with the best fit solution located at the low-$\Gamma$ end. Changing the injection's power-law index away from the value $p = 2.5$ also preserves the topology, but the location of the best fit solution shifts somewhat from the low-$\Gamma$ end of the valley (see Figs.~\ref{HardParameterMap} and \ref{SoftParameterMap}) 
and goodness of fit is worse. Finally, we find the same topology  for two moments of time that we explored --- at $t_\mathrm{obs} = 90$~s and at $t_\mathrm{obs} = 145$~s. 

We  {analyzed} each of the two SEDs separately, 
namely
we  determined the conditions at the early and late time independently of each other.
It turns out that both for the early and late  time  the best fit SEDs correspond to fast cooling solutions with significant self-absorption of IC photons. 
Both solutions have relatively large magnetization and energy fraction in accelerated electrons ($\epsB \sim 0.003 \div 0.006$ and $\epse \sim 0.1$ assuming $E_\mathrm{kin} = 3 \times 10^{53}$~erg). 
Their ratio, $\Ksc \sim 20 \div 40$, agrees with common expectations, but our fits suggest it increases in time.

The early-time and late-time solutions are consistent with constant $\epse$, but, surprisingly, are not consistent with constant $\epsB$ under assumption of adiabatic shock. The latter inconsistency suggests $\gtrsim 20\%$ decrease of shock's kinetic energy between $t_\mathrm{obs} = 90$~s and $t_\mathrm{obs} = 145$~s. 
These radiative losses are reasonable at  such an early phase of the afterglow. 
Parameters of both fits approximately satisfy the relation $\langle \gamma \rangle \simeq \left( B_\mathrm{cr}/B \right)^{1/3}$. We also find that the  fraction of self-absorbed photons stays  approximately constant. Both features are predictions of the pair-balance model but are coincidental in others. More surprising is the fact that the fits at two moments of time also demonstrate that the average Lorentz factor of injected electrons increases in time  whereas the shock's Lorentz factor decreases. This is counter to the common expectations, but finds explanation within the pair-balance model. 

The parameters of our best fit solutions that we find under assumption of constant shock's kinetic energy are summarized in Table~\ref{ParametersSummary}. The exact  values we obtain depend on the set of {model} coefficients that we used and on the  smooth shape of the injection function that we have assumed. 
This treatment is meant to give us better estimate of the conditions in the emitting region. By no means the specific set of coefficients and specific choice of the injection function's shape is the reason for  deviation from the common model --- the main conclusions of incompatibility with the common assumptions 
are independent
of these choices and they hold, for example, for \SPN{} and a truncated injection function.

\section{Summary}
\label{sec:claims}

In this work we have used a new code \citep{newcode} to calculate the SSC spectrum and obtain a best fit to the observations of GRB~190114C. The calculations include exact  KN  corrections, pair production via self absorption of the high energy photons and the corresponding emission of these pairs. Both the IC losses and the production of secondaries are calculated via kinetic equations with exact QED cross-sections.
This modified one-zone model introduces a new set of coefficients that enable us to mimic within a single zone calculations the effect of a quasi-spherical {relativistic} blast wave. 

Our most important findings are:
\begin{itemize} 
\item  The best-fit solutions to the observed GRB~190114C SEDs both at $t_\mathrm{obs} = 90$~s and at $t_\mathrm{obs} = 145$~s are fast cooling (only $\sim 10\%$ of the power 
of the injected electrons'  is not radiated) with  KN  corrections to the inverse Compton spectra and  relatively strong self-absorption of IC photons 
{($\simeq 10$\% of the total emitted power, i.e., $\simeq 25$\% of initially produced IC power, is absorbed in situ).}
They provide good fits to the observed SEDs from  optical to TeV despite they were obtained without reference to the optical flux measurements (i.e., we find no need in an additional source of optical radiation).
\item We find that afterglow blast wave of GRB~190114C had an isotropic equivalent energy of at least $1.3 \times 10^{53}$~erg 
(and likely close to this lower limit).
The afterglow  blast wave was still highly relativistic $\Gamma \gtrsim 160$ 
at 90~s after the burst. A lower limit $\Gamma >100$ was derived by \cite{DerishevPiran2019} ignoring optical emission from secondary pairs. 
The  equipartition parameters are 
$\epse \simeq 0.12 / E_{53.5}$, $\epsB \simeq 0.006 / E_{53.5}$ at $t_\mathrm{obs}=90$~s and $\epse \simeq 0.1 / E_{53.5}$, $\epsB \simeq 0.0027 / E_{53.5}$ at $t_\mathrm{obs}=145$~s.
{The} fit parameters at  the two moments of time are consistent with constant $\epse$ {(with $\sim 20\%$ decrease in shock's kinetic energy)} but not with  constant $\epsB$. 
\item 
{The best fit numerical solution that we find has characteristics similar to analytic estimate by \cite{DerishevPiran2019} --- it is fast cooling with significant self-absorption and Comptonization on the border between Thomson and  KN  regimes.}
\item The values for the {best-fit} parameters we obtain are different from those found in earlier works
\citep{MAGIC_Nature_fit,Wang_etal_SED,Asano_etal_SED}. In particular we cannot reproduce the theoretical spectrum shown in the discovery paper \citep{MAGIC_Nature_fit} with their given parameters; we reproduce the TeV part shown in \cite{Wang_etal_SED} however, this work doesn't attempt to fit both X-rays and TeV {with a single} source; finally we reproduce the spectrum presented by \cite{Asano_etal_SED} once we adopt their relation between the escape time of the electrons and the photons. However, {their} solution is {substantially} different from {the best fit that we find.} 
\item The typical Lorentz factor of injected electrons, $\gamma_\mathrm{b}$ or $\gamma_\mathrm{m}$, approximately satisfies the prediction of pair-balance model that it should be $\sim \left( B_\mathrm{cr} / B \right)^{1/3}$, and it increases in time.  Such increase is predicted in the pair-balance model, but it contradicts the common assumption that  
{$\gamma_\mathrm{b}, \gamma_\mathrm{m} \propto \Gamma$}.
\item  
The TeV data points are the most important factor in determining the acceptable range for $B$, $\Ksc$, and $\gamma_\mathrm{b}$. In combination with the GeV data they set also a lower limit $\Gamma \gtrsim 140$. However, the optical flux measurements put even more stringent lower limit and also an upper limit: 
$160 < \Gamma < 430$.  {The best fit solution 
that also matches the observed optical flux, 
has $\Gamma \simeq 160$  coinciding with the lower limit.}
\end{itemize}

To conclude we note that while GRB~190114C is the first GRB detected in TeV, there is no reason to expect that it is unique. The observed deviation  from common modeling assumptions may be generic, and if so, these observations suggest that we have to reconsider  these assumptions.
Additional  detections of TeV emission from GRBs with 
Cherenkov telescopes and in particular the upcoming CTA may  demonstrate this need in the future.

\section{Acknowledgment}
We thank Ehud Nakar, Lara Nava, Alexei Pozanenko and Shotaro Yamasaki for helpful discussions. This research is supported by the Russian Science Foundation grant No. 21-12-00416 (ED) and by an advanced ERC grant: TREX (TP).


\begin{thebibliography}{}
\expandafter\ifx\csname natexlab\endcsname\relax\def\natexlab#1{#1}\fi
\providecommand{\url}[1]{\href{#1}{#1}}
\providecommand{\dodoi}[1]{doi:~\href{http://doi.org/#1}{\nolinkurl{#1}}}
\providecommand{\doeprint}[1]{\href{http://ascl.net/#1}{\nolinkurl{http://ascl.net/#1}}}
\providecommand{\doarXiv}[1]{\href{https://arxiv.org/abs/#1}{\nolinkurl{https://arxiv.org/abs/#1}}}

\bibitem[{{Aksulu} {et~al.}(2020){Aksulu}, {Wijers}, {van Eerten}, \& {van der
  Horst}}]{Aksulu2020}
{Aksulu}, M.~D., {Wijers}, R.~A.~M.~J., {van Eerten}, H.~J., \& {van der
  Horst}, A.~J. 2020, \mnras, 497, 4672, \dodoi{10.1093/mnras/staa2297}

\bibitem[{{Ando} {et~al.}(2008){Ando}, {Nakar}, \& {Sari}}]{AndoNakarSari2008}
{Ando}, S., {Nakar}, E., \& {Sari}, R. 2008, \apj, 689, 1150,
  \dodoi{10.1086/592486}

\bibitem[{{Asano} {et~al.}(2020){Asano}, {Murase}, \& {Toma}}]{Asano_etal_SED}
{Asano}, K., {Murase}, K., \& {Toma}, K. 2020, \apj, 905, 105,
  \dodoi{10.3847/1538-4357/abc82c}

\bibitem[{{Barniol Duran} {et~al.}(2012){Barniol Duran}, {Bo{\v{s}}njak}, \&
  {Kumar}}]{KNeffects_BarniolDuranBosnjakKumar}
{Barniol Duran}, R., {Bo{\v{s}}njak}, {\v{Z}}., \& {Kumar}, P. 2012, \mnras,
  424, 3192, \dodoi{10.1111/j.1365-2966.2012.21533.x}

\bibitem[{{Beniamini} {et~al.}(2015){Beniamini}, {Nava}, {Duran}, \&
  {Piran}}]{Beniamini2015}
{Beniamini}, P., {Nava}, L., {Duran}, R.~B., \& {Piran}, T. 2015, \mnras, 454,
  1073, \dodoi{10.1093/mnras/stv2033}

\bibitem[{{Beniamini} \& {Piran}(2014)}]{BeniaminiPiran2014}
{Beniamini}, P., \& {Piran}, T. 2014, \mnras, 445, 3892,
  \dodoi{10.1093/mnras/stu2032}

\bibitem[{{Blandford} \& {McKee}(1976)}]{BlandfordMcKee}
{Blandford}, R.~D., \& {McKee}, C.~F. 1976, Physics of Fluids, 19, 1130,
  \dodoi{10.1063/1.861619}

\bibitem[{{Dai} \& {Lu}(1998)}]{DaiLu98}
{Dai}, Z.~G., \& {Lu}, T. 1998, \mnras, 298, 87,
  \dodoi{10.1046/j.1365-8711.1998.01681.x}

\bibitem[{{Daigne} {et~al.}(2011){Daigne}, {Bo{\v{s}}njak}, \&
  {Dubus}}]{KNeffects_DaigneBosnjakDubus}
{Daigne}, F., {Bo{\v{s}}njak}, {\v{Z}}., \& {Dubus}, G. 2011, \aap, 526, A110,
  \dodoi{10.1051/0004-6361/201015457}

\bibitem[{{Derishev}(2021)}]{newcode}
{Derishev}, E. 2021, to be submitted

\bibitem[{{Derishev} \& {Piran}(2019)}]{DerishevPiran2019}
{Derishev}, E., \& {Piran}, T. 2019, \apjl, 880, L27,
  \dodoi{10.3847/2041-8213/ab2d8a}

\bibitem[{{Derishev} {et~al.}(2003){Derishev}, {Aharonian}, {Kocharovsky}, \&
  {Kocharovsky}}]{converter}
{Derishev}, E.~V., {Aharonian}, F.~A., {Kocharovsky}, V.~V., \& {Kocharovsky},
  V.~V. 2003, \prd, 68, 043003, \dodoi{10.1103/PhysRevD.68.043003}

\bibitem[{{Derishev} {et~al.}(2001){Derishev}, {Kocharovsky}, \&
  {Kocharovsky}}]{DerishevKocharovskyKocharovsky2001}
{Derishev}, E.~V., {Kocharovsky}, V.~V., \& {Kocharovsky}, V.~V. 2001, \aap,
  372, 1071, \dodoi{10.1051/0004-6361:20010586}

\bibitem[{{Derishev} \& {Piran}(2016)}]{PairBalance}
{Derishev}, E.~V., \& {Piran}, T. 2016, \mnras, 460, 2036,
  \dodoi{10.1093/mnras/stw1175}

\bibitem[{{Fan}(2010)}]{KNeffects_Fan}
{Fan}, Y.-Z. 2010, \mnras, 403, 483, \dodoi{10.1111/j.1365-2966.2009.16134.x}

\bibitem[{{Fan} {et~al.}(2008){Fan}, {Piran}, {Narayan}, \&
  {Wei}}]{FanPiranNarayan2008}
{Fan}, Y.-Z., {Piran}, T., {Narayan}, R., \& {Wei}, D.-M. 2008, \mnras, 384,
  1483, \dodoi{10.1111/j.1365-2966.2007.12765.x}

\bibitem[{{Fitzpatrick}(1999)}]{ExtinctionModel}
{Fitzpatrick}, E.~L. 1999, \pasp, 111, 63, \dodoi{10.1086/316293}

\bibitem[{{Ghirlanda} {et~al.}(2018){Ghirlanda}, {Nappo}, {Ghisellini},
  {Melandri}, {Marcarini}, {Nava}, {Salafia}, {Campana}, \&
  {Salvaterra}}]{Ghirlanda+2018}
{Ghirlanda}, G., {Nappo}, F., {Ghisellini}, G., {et~al.} 2018, \aap, 609, A112,
  \dodoi{10.1051/0004-6361/201731598}

\bibitem[{{Granot} \& {Sari}(2002)}]{GranotSari}
{Granot}, J., \& {Sari}, R. 2002, \apj, 568, 820, \dodoi{10.1086/338966}

\bibitem[{{Lemoine}(2015)}]{KNeffects_Lemoine}
{Lemoine}, M. 2015, \mnras, 453, 3772, \dodoi{10.1093/mnras/stv1800}

\bibitem[{{MAGIC Collaboration} {et~al.}(2019{\natexlab{a}}){MAGIC
  Collaboration}, {Acciari}, {Ansoldi}, {Antonelli}, {Arbet Engels}, {Baack},
  {Babi{\'c}}, {Banerjee}, {Barres de Almeida}, {Barrio}, {Becerra
  Gonz{\'a}lez}, {Bednarek}, {Bellizzi}, {Bernardini}, {Berti}, {Besenrieder},
  {Bhattacharyya}, {Bigongiari}, {Biland }, {Blanch}, {Bonnoli},
  {Bo{\v{s}}njak}, {Busetto}, {Carosi}, {Carosi}, {Ceribella}, {Chai},
  {Chilingaryan}, {Cikota}, {Colak}, {Colin}, {Colombo}, {Contreras},
  {Cortina}, {Covino}, {D'Amico}, {D'Elia}, {da Vela}, {Dazzi}, {de Angelis},
  {de Lotto}, {Delfino}, {Delgado}, {Depaoli}, {di Pierro}, {di Venere}, {Do
  Souto Espi{\~n}eira}, {Dominis Prester}, {Donini}, {Dorner}, {Doro},
  {Elsaesser}, {Fallah Ramazani}, {Fattorini}, {Fern{\'a}ndez-Barral},
  {Ferrara}, {Fidalgo}, {Foffano}, {Fonseca}, {Font}, {Fruck}, {Fukami},
  {Gallozzi}, {Garc{\'\i}a L{\'o}pez}, {Garczarczyk}, {Gasparyan}, {Gaug},
  {Giglietto}, {Giordano}, {Godinovi{\'c}}, {Green}, {Guberman}, {Hadasch},
  {Hahn}, {Herrera}, {Hoang}, {Hrupec}, {H{\"u}tten}, {Inada}, {Inoue},
  {Ishio}, {Iwamura}, {Jouvin}, {Kerszberg}, {Kubo}, {Kushida}, {Lamastra},
  {Lelas}, {Leone}, {Lindfors}, {Lombardi}, {Longo}, {L{\'o}pez},
  {L{\'o}pez-Coto}, {L{\'o}pez-Oramas}, {Loporchio}, {Machado de Oliveira
  Fraga}, {Maggio}, {Majumdar}, {Makariev}, {Mallamaci}, {Maneva}, {Manganaro},
  {Mannheim}, {Maraschi}, {Mariotti}, {Mart{\'\i}nez}, {Masuda}, {Mazin},
  {Mi{\'c}anovi{\'c}}, {Miceli}, {Minev}, {Miranda}, {Mirzoyan}, {Molina},
  {Moralejo}, {Morcuende}, {Moreno}, {Moretti}, {Munar-Adrover}, {Neustroev},
  {Nigro}, {Nilsson}, {Ninci}, {Nishijima}, {Noda}, {Nogu{\'e}s}, {N{\"o}the},
  {Nozaki}, {Paiano}, {Palacio}, {Palatiello}, {Paneque}, {Paoletti},
  {Paredes}, {Pe{\~n}il}, {Peresano}, {Persic}, {Prada Moroni}, {Prand ini},
  {Puljak}, {Rhode}, {Rib{\'o}}, {Rico}, {Righi}, {Rugliancich}, {Saha},
  {Sahakyan}, {Saito}, {Sakurai}, {Satalecka}, {Schmidt}, {Schweizer},
  {Sitarek}, {{\v{S}}nidari{\'c}}, {Sobczynska}, {Somero}, {Stamerra}, {Strom},
  {Strzys}, {Suda}, {Suri{\'c}}, {Takahashi}, {Tavecchio}, {Temnikov},
  {Terzi{\'c}}, {Teshima}, {Torres-Alb{\`a}}, {Tosti}, {Tsujimoto}, {Vagelli},
  {van Scherpenberg}, {Vanzo}, {Vazquez Acosta}, {Vigorito}, {Vitale}, {Vovk},
  {Will}, {Zari{\'c}}, \& {Nava}}]{MAGIC_Nature_obs}
{MAGIC Collaboration}, {Acciari}, V.~A., {Ansoldi}, S., {et~al.}
  2019{\natexlab{a}}, \nat, 575, 455, \dodoi{10.1038/s41586-019-1750-x}

\bibitem[{{MAGIC Collaboration} {et~al.}(2019{\natexlab{b}}){MAGIC
  Collaboration}, {Acciari}, {Ansoldi}, {Antonelli}, {Engels}, {Baack},
  {Babi{\'c}}, {Banerjee}, {Barres de Almeida}, {Barrio}, {Becerra
  Gonz{\'a}lez}, {Bednarek}, {Bellizzi}, {Bernardini}, {Berti}, {Besenrieder},
  {Bhattacharyya}, {Bigongiari}, {Biland }, {Blanch}, {Bonnoli},
  {Bo{\v{s}}njak}, {Busetto}, {Carosi}, {Ceribella}, {Chai}, {Chilingaryan},
  {Cikota}, {Colak}, {Colin}, {Colombo}, {Contreras}, {Cortina}, {Covino},
  {D'Elia}, {da Vela}, {Dazzi}, {de Angelis}, {de Lotto}, {Delfino}, {Delgado},
  {Depaoli}, {di Pierro}, {di Venere}, {Do Souto Espi{\~n}eira}, {Dominis
  Prester}, {Donini}, {Dorner}, {Doro}, {Elsaesser}, {Fallah Ramazani},
  {Fattorini}, {Ferrara}, {Fidalgo}, {Foffano}, {Fonseca}, {Font}, {Fruck},
  {Fukami}, {Garc{\'\i}a L{\'o}pez}, {Garczarczyk}, {Gasparyan}, {Gaug},
  {Giglietto}, {Giordano}, {Godinovi{\'c}}, {Green}, {Guberman}, {Hadasch},
  {Hahn}, {Herrera}, {Hoang}, {Hrupec}, {H{\"u}tten}, {Inada}, {Inoue},
  {Ishio}, {Iwamura}, {Jouvin}, {Kerszberg}, {Kubo}, {Kushida}, {Lamastra},
  {Lelas}, {Leone}, {Lindfors}, {Lombardi}, {Longo}, {L{\'o}pez},
  {L{\'o}pez-Coto}, {L{\'o}pez-Oramas}, {Loporchio}, {Machado de Oliveira
  Fraga}, {Maggio}, {Majumdar}, {Makariev}, {Mallamaci}, {Maneva}, {Manganaro},
  {Mannheim}, {Maraschi}, {Mariotti}, {Mart{\'\i}nez}, {Mazin},
  {Mi{\'c}anovi{\'c}}, {Miceli}, {Minev}, {Mirand a}, {Mirzoyan}, {Molina},
  {Moralejo}, {Morcuende}, {Moreno}, {Moretti}, {Munar-Adrover}, {Neustroev},
  {Nigro}, {Nilsson}, {Ninci}, {Nishijima}, {Noda}, {Nogu{\'e}s}, {Nozaki},
  {Paiano}, {Palatiello}, {Paneque}, {Paoletti}, {Paredes}, {Pe{\~n}il},
  {Peresano}, {Persic}, {Moroni}, {Prandini}, {Puljak}, {Rhode}, {Rib{\'o}},
  {Rico}, {Righi}, {Rugliancich}, {Saha}, {Sahakyan}, {Saito}, {Sakurai},
  {Satalecka}, {Schmidt}, {Schweizer}, {Sitarek}, {{\v{S}}nidari{\'c}},
  {Sobczynska}, {Somero}, {Stamerra}, {Strom}, {Strzys}, {Suda}, {Suri{\'c}},
  {Takahashi}, {Tavecchio}, {Temnikov}, {Terzi{\'c}}, {Teshima},
  {Torres-Alb{\`a}}, {Tosti}, {Vagelli}, {van Scherpenberg}, {Vanzo}, {Vazquez
  Acosta}, {Vigorito}, {Vitale}, {Vovk}, {Will}, {Zari{\'c}}, {Nava}, {Veres},
  {Bhat}, {Briggs}, {Cleveland }, {Hamburg}, {Hui}, {Mailyan}, {Preece},
  {Roberts}, {von Kienlin}, {Wilson-Hodge}, {Kocevski}, {Arimoto}, {Tak},
  {Asano}, {Axelsson}, {Barbiellini}, {Bissaldi}, {Dirirsa}, {Gill}, {Granot},
  {McEnery}, {Omodei}, {Razzaque}, {Piron}, {Racusin}, {Thompson}, {Campana},
  {Bernardini}, {Kuin}, {Siegel}, {Cenko}, {O'Brien}, {Capalbi}, {Da{\i}}, {de
  Pasquale}, {Gropp}, {Klingler}, {Osborne}, {Perri}, {Starling},
  {Tagliaferri}, {Tohuvavohu}, {Ursi}, {Tavani}, {Cardillo}, {Casentini},
  {Piano}, {Evangelista}, {Verrecchia}, {Pittori}, {Lucarelli}, {Bulgarelli},
  {Parmiggiani}, {Anderson}, {Anderson}, {Bernardi}, {Bolmer},
  {Caballero-Garc{\'\i}a}, {Carrasco}, {Castell{\'o}n}, {Castro Segura},
  {Castro-Tirado}, {Cherukuri}, {Cockeram}, {D'Avanzo}, {di Dato}, {Diretse},
  {Fender}, {Fern{\'a}ndez-Garc{\'\i}a}, {Fynbo}, {Fruchter}, {Greiner},
  {Gromadzki}, {Heintz}, {Heywood}, {van der Horst}, {Hu}, {Inserra}, {Izzo},
  {Jaiswal}, {Jakobsson}, {Japelj}, {Kankare}, {Kann}, {Kouveliotou}, {Klose},
  {Levan}, {Li}, {Lotti}, {Maguire}, {Malesani}, {Manulis}, {Marongiu},
  {Martin}, {Melandri}, {Micha{\l}owski}, {Miller-Jones}, {Misra}, {Moin},
  {Mooley}, {Nasri}, {Nicholl}, {Noschese}, {Novara}, {Pandey}, {Peretti},
  {P{\'e}rez Del Pulgar}, {P{\'e}rez-Torres}, {Perley}, {Piro}, {Ragosta},
  {Resmi}, {Ricci}, {Rossi}, {S{\'a}nchez-Ram{\'\i}rez}, {Selsing}, {Schulze},
  {Smartt}, {Smith}, {Sokolov}, {Stevens}, {Tanvir}, {Th{\"o}ne}, {Tiengo},
  {Tremou}, {Troja}, {de Ugarte Postigo}, {Valeev}, {Vergani}, {Wieringa},
  {Woudt}, {Xu}, {Yaron}, \& {Young}}]{MAGIC_Nature_fit}
---. 2019{\natexlab{b}}, \nat, 575, 459, \dodoi{10.1038/s41586-019-1754-6}

\bibitem[{{M{\'e}sz{\'a}ros} \& {Rees}(1997)}]{MeszarosRees97}
{M{\'e}sz{\'a}ros}, P., \& {Rees}, M.~J. 1997, \apj, 476, 232,
  \dodoi{10.1086/303625}

\bibitem[{{Nakar} {et~al.}(2009){Nakar}, {Ando}, \& {Sari}}]{NakarAndoSari2009}
{Nakar}, E., {Ando}, S., \& {Sari}, R. 2009, \apj, 703, 675,
  \dodoi{10.1088/0004-637X/703/1/675}

\bibitem[{{Nava} {et~al.}(2013){Nava}, {Sironi}, {Ghisellini}, {Celotti}, \&
  {Ghirlanda}}]{Nava+13}
{Nava}, L., {Sironi}, L., {Ghisellini}, G., {Celotti}, A., \& {Ghirlanda}, G.
  2013, \mnras, 433, 2107, \dodoi{10.1093/mnras/stt872}

\bibitem[{{Panaitescu} \& {Kumar}(2000)}]{PanaitescuKumar2000}
{Panaitescu}, A., \& {Kumar}, P. 2000, \apj, 543, 66, \dodoi{10.1086/317090}

\bibitem[{{Panaitescu} \& {Kumar}(2002)}]{PanaitescKumar2002}
---. 2002, \apj, 571, 779, \dodoi{10.1086/340094}

\bibitem[{{Panaitescu} \&
  {M{\'e}sz{\'a}ros}(1998{\natexlab{a}})}]{PanaitescuMeszaros98}
{Panaitescu}, A., \& {M{\'e}sz{\'a}ros}, P. 1998{\natexlab{a}}, \apj, 501, 772,
  \dodoi{10.1086/305856}

\bibitem[{{Panaitescu} \&
  {M{\'e}sz{\'a}ros}(1998{\natexlab{b}})}]{PanaitescuMeszaros98b}
---. 1998{\natexlab{b}}, \apjl, 493, L31, \dodoi{10.1086/311127}

\bibitem[{{Pe'er} \& {Waxman}(2005)}]{PeerWaxman2005}
{Pe'er}, A., \& {Waxman}, E. 2005, \apj, 633, 1018, \dodoi{10.1086/468175}

\bibitem[{{Sari}(1997)}]{Sari97}
{Sari}, R. 1997, \apjl, 489, L37, \dodoi{10.1086/310957}

\bibitem[{{Sari} \& {Esin}(2001)}]{SariEsin2001}
{Sari}, R., \& {Esin}, A.~A. 2001, \apj, 548, 787, \dodoi{10.1086/319003}

\bibitem[{{Sari} {et~al.}(1996){Sari}, {Narayan}, \&
  {Piran}}]{SariNarayanPiran96}
{Sari}, R., {Narayan}, R., \& {Piran}, T. 1996, \apj, 473, 204,
  \dodoi{10.1086/178136}

\bibitem[{{Sari} {et~al.}(1998){Sari}, {Piran}, \&
  {Narayan}}]{AfterglowModelling}
{Sari}, R., {Piran}, T., \& {Narayan}, R. 1998, \apjl, 497, L17,
  \dodoi{10.1086/311269}

\bibitem[{{Schlafly} \& {Finkbeiner}(2011)}]{GalacticExtinction}
{Schlafly}, E.~F., \& {Finkbeiner}, D.~P. 2011, \apj, 737, 103,
  \dodoi{10.1088/0004-637X/737/2/103}

\bibitem[{{Tak} {et~al.}(2019){Tak}, {Omodei}, {Uhm}, {Racusin}, {Asano}, \&
  {McEnery}}]{KNeffects_TakOmodeiUhm}
{Tak}, D., {Omodei}, N., {Uhm}, Z.~L., {et~al.} 2019, \apj, 883, 134,
  \dodoi{10.3847/1538-4357/ab3982}

\bibitem[{{van Eerten} {et~al.}(2012){van Eerten}, {van der Horst}, \&
  {MacFadyen}}]{vanEerte+2012}
{van Eerten}, H., {van der Horst}, A., \& {MacFadyen}, A. 2012, \apj, 749, 44,
  \dodoi{10.1088/0004-637X/749/1/44}

\bibitem[{{Wang} {et~al.}(2010){Wang}, {He}, {Li}, {Wu}, \&
  {Dai}}]{KNeffects_WangHeLi}
{Wang}, X.-Y., {He}, H.-N., {Li}, Z., {Wu}, X.-F., \& {Dai}, Z.-G. 2010, \apj,
  712, 1232, \dodoi{10.1088/0004-637X/712/2/1232}

\bibitem[{{Wang} {et~al.}(2019){Wang}, {Liu}, {Zhang}, {Xi}, \&
  {Zhang}}]{Wang_etal_SED}
{Wang}, X.-Y., {Liu}, R.-Y., {Zhang}, H.-M., {Xi}, S.-Q., \& {Zhang}, B. 2019,
  \apj, 884, 117, \dodoi{10.3847/1538-4357/ab426c}

\bibitem[{{Waxman}(1997)}]{Waxman97}
{Waxman}, E. 1997, \apjl, 491, L19, \dodoi{10.1086/311057}

\bibitem[{{Wijers} \& {Galama}(1999)}]{WijersGalama1999}
{Wijers}, R.~A.~M.~J., \& {Galama}, T.~J. 1999, \apj, 523, 177,
  \dodoi{10.1086/307705}

\bibitem[{{Yamasaki} \& {Piran}(2021)}]{Yamasaki+21}
{Yamasaki}, S., \& {Piran}, T. 2021, to be submitted

\end{thebibliography}
\end{document}